\documentclass[fleqn,usenatbib]{mnras}

\usepackage[T1]{fontenc}

\DeclareRobustCommand{\VAN}[3]{#2}
\let\VANthebibliography\thebibliography
\def\thebibliography{\DeclareRobustCommand{\VAN}[3]{##3}\VANthebibliography}

\usepackage{graphicx}	
\usepackage{amsmath}	
\usepackage{amssymb}	

\usepackage{newtxtext,newtxmath}

\title[Period Change Rates of LMC Cepheids using MESA]{Period Change Rates of Large Magellanic Cloud Cepheids using MESA}

\author[Espinoza-Arancibia et al.]{
F. Espinoza-Arancibia,$^{1,2,3}$\thanks{E-mail: fespinoza@camk.edu.pl}
M. Catelan,$^{1,2}$
G. Hajdu,$^{3}$
N. Rodr\'{i}guez-Segovia,$^{1,4}$
G. Boggiano,$^{1}$
\newauthor
K. Joachimi,$^{1}$
C. Mu\~{n}oz-L\'{o}pez,$^{1}$
C. Ordenes-Huanca,$^{1,2}$
C. Orquera-Rojas$,^{1,2}$
P. Torres$,^{1,2}$
\newauthor
\'{A}. Valenzuela-Navarro$^{1,2}$
\\
$^{1}$Instituto de Astrof\'{i}sica, Facultad de Física, Pontificia Universidad Cat\'{o}lica de Chile, Av. Vicu\~{n}a Mackenna 4860, 7820436 Macul, Santiago, Chile\\
$^{2}$Millennium Institute of Astrophysics, Nuncio Monseñor Sotero Sanz 100, Of. 104, Providencia, Santiago, Chile\\
$^{3}$Nicolaus Copernicus Astronomical Center, Polish Academy of Sciences, Bartycka 18, 00-716 Warsaw, Poland\\
$^{4}$School of Science, University of New South Wales, Australian Defence Force Academy, Canberra, ACT 2600, Australia
}

\date{Accepted XXX. Received YYY; in original form ZZZ}

\pubyear{2022}

\begin{document}
\label{firstpage}
\pagerange{\pageref{firstpage}--\pageref{lastpage}}
\maketitle

\begin{abstract}
Pulsating stars, such as Cepheids and RR Lyrae, offer us a window to measure and study changes due to stellar evolution. In this work, we study the former by calculating a set of evolutionary tracks of stars with an initial mass of 4 to 7 $M_{\sun}$, varying the initial rotation rate and metallicity, using the stellar evolution code Modules for Experiments in Stellar Astrophysics (MESA). Using Radial Stellar Pulsations (RSP), a recently added functionality of MESA, we obtained theoretical instability strip (IS) edges and linear periods for the radial fundamental mode. Period-age, period-age-temperature, period-luminosity, and period-luminosity-temperature relationships were derived for three rotation rates and metallicities, showing a dependence on crossing number, position in the IS, rotation, and metallicity. We calculated period change rates (PCRs) based on the linear periods from RSP. We compared our models with literature results using the Geneva code, and found large differences, as expected due to the different implementations of rotation between codes. In addition, we compared our theoretical PCRs with those measured in our recent work for Large Magellanic Cloud Cepheids. We found good overall agreement, even though our models do not reach the short-period regime exhibited by the empirical data. Implementations of physical processes not yet included in our models, such as pulsation-driven mass loss, an improved treatment of convection that may lead to a better description of the instability strip edges, as well as consideration of a wider initial mass range, could all help improve the agreement with the observed PCRs.
\end{abstract}


\section{Introduction}

Classical Cepheids, also known as Type I Cepheids (hereafter Cepheids), are radially pulsating variable stars located in the instability strip (IS) of the Hertzsprung-Russel diagram (HRD). As evolved intermediate-mass stars, they typically have between 2 and 13 ${\rm M}_{\odot}$ \citep[see, e.g.][]{Bono2000,Anderson2016}, though Cepheids with even higher masses are also known \citep[][and references therein]{Musella2022}. Such stars first cross the instability strip, and become Cepheids, after they depart from the main sequence (MS) and cross the Hertzsprung gap, where core H-burning ceases and H-shell burning dominates. Later, the star crosses the IS again (at least twice) during the blue loop, an evolutionary phase during which the star fuses helium in its core. The time scale of the first crossing is over 10 times faster than that of subsequent crossings, and thus almost all observed Cepheids can be considered to evolve along the blue loop. The pulsations of these stars are driven by mechanisms that operate in the regions where we find partially ionized H and He~-- the so-called $\kappa$ and $\gamma$ mechanisms \citep[see, e.g.,][and references therein]{catelan2015}. 

\par Although already past the main sequence, Cepheids are young stars, with ages ranging between $\sim10^7$ years for the massive, more luminous Cepheids to $\sim10^8$ years for the least-massive, fainter ones. For this reason, Cepheids are only found in regions of recent star formation. In the Milky Way, they are located in the young disk, while they are also identified in nearby galaxies such as the Andromeda Galaxy (M31) and the Magellanic Clouds \citep[][and references therein]{catelan2015}. Since Cepheids obey a tight relation between their pulsation period and luminosity~-- the so-called {\em period-luminosity relation} or {\em Leavitt law} \citep{leavitt1908,leavitt1912}~-- they are standard candles for determining distances to nearby galaxies, making them crucial objects in many determinations of the Hubble constant \citep[e.g.,][]{riess2019,2021ApJ...911...12J}.

Over the last decades, there has been a large increase in the number of surveys and amount of data available for the study of variable stars. Two such surveys, namely the Massive Compact Halo Objects  \citep[MACHO;][]{alcock1996} and Optical Gravitational Lensing Experiment \citep[OGLE;][]{1994IAUC.5997....1P} surveys, both of which were originally designed to search for gravitational microlensing events, observed (and, in the OGLE case, continue to observe) continuously thousands of Cepheids in the Magellanic Clouds and Milky Way bulge. \cite{Soszynski2015} presented an almost complete census of classical Cepheids in the Magellanic Clouds, consisting of 9535 Cepheids, of which 4620 belong to the Large Magellanic Cloud (LMC) and 4915 are members of the Small Magellanic Cloud (SMC). In like vein, \cite{2021AcA....71..205P} presented a collection of 3352 Galactic classical Cepheids. In recent years, and especially with the advent of its second and third data releases (DR2 and DR3, respectively), \textit{Gaia} is playing an especially important role in characterizing nearby Cepheids \citep{Clementini2019,Rimoldini2019,Ripepi2022}, and indeed other types of pulsating variables
\citep{Rimoldini2019,Eyer2022,Clementini2022,DeRidder2022,Lebzelter2022}. In particular, the 34 months of data contained in \textit{Gaia}'s DR3 has led to multi-band time series photometry for 4663, 4616, 321, and 185 Cepheids located in the LMC, SMC, M31, and M33, respectively, as well as 5221 objects in the remaining fields. Recent progress in spectroscopic studies of classical Cepheids are also worth noting \citep[e.g.,][see also \citeauthor{Groenewegen2018} \citeyear{Groenewegen2018}, and references therein]{Ripepi2021,daSilva2022}. In addition to this abundance of recent data, a source of historical light curves for long-term variability studies is the Digital Access to a Sky Century\,@\,Harvard \citep[DASCH; ][]{grindlay2011} project, which aims to digitize the majority of the Harvard College Observatory's Astronomical Photographic Plate Collection (HCOAPPC), and provide photometry based on these plates.

\par With this enormous amount of long-time baseline data, period change rates (PCRs) of different types of pulsating stars have been measured, and the results compared with theoretical predictions. \cite{Turner2006} used observations from HCOAPPC to measure period changes in nearly 200 Galactic Cepheids. They found that about two-thirds of those have positive PCRs, whereas the remaining have negative rates, in broad agreement with contemporary stellar evolution models. \cite{Neilson2012} constructed state-of-the-art evolutionary Cepheid models that considered enhanced mass loss compared to canonical stellar evolution models, and reached results consistent with those of \cite{Turner2006}. \cite{Anderson2014,Anderson2016} calculated new stellar evolution models that include rotation using the Geneva code. They showed that rotation has a strong impact on the evolution of Cepheids, and that including rotation in the models may be necessary to understand the observational PCRs. Recently, \cite{Miller2020} concluded that stellar rotation together with convective core overshooting are not sufficient to explain the empirical PCRs, proposing pulsation-driven mass loss as a mechanism that should be additionally considered.

\par In this work, we use version 11701 of \textsc{Modules for Experiments in Stellar Astrophysics
}\citep[MESA; ][]{Paxton2011,Paxton2013,Paxton2015,Paxton2018,Paxton2019}, a state-of-the-art, open-source 1D stellar evolution code, to compute a grid of evolutionary tracks of intermediate-mass stars, including rotation, covering from the MS to the end of the blue loop phase. With Radial Stellar Pulsation (RSP), a new functionality of MESA \citep{Smolec2008,Paxton2019}, we also computed PCRs for the three IS crossings presented by each evolutionary track. In addition, we studied the effect of rotation on the PCR for these stars, and compared the results with those obtained by \cite{Anderson2016}. In our previous paper \citep{Rodriguez2021}, we derived PCRs for classical LMC Cepheids using DASCH, OGLE, and other datasets. Through a comparison with our model results, in this paper we discuss the evolutionary status of the stars in the \citep{Rodriguez2021} sample.

\par The outline of this paper is as follows: Section~\ref{sec:adoptedphysics} describes the physics adopted in our MESA models, adopted parameters, and calibrations; Section~\ref{sec:Models} describes the main features of the obtained evolutionary models, discussing their implications and comparing them with models from \citet{Anderson2016}; Section~\ref{sec:pdotMESA} presents the periods and PCRs obtained using the RSP functionality, in addition to period-age and period-luminosity relationships, as well as a comparison with PCRs from other works. Finally, Section~\ref{sec:conclusions} presents our conclusions and future directions.

\section{Adopted physics} \label{sec:adoptedphysics}
\subsection{Abundances}\label{sec:maths}
In this paper, we adopt solar-scaled abundances, based on a solar mix as provided by \cite{Grevesse1998}. Our choice is motivated by the fact that these abundances provide a better match to helioseismological constraints than do those based on the 3D hydrodynamical analyses by \citet{Asplund2005,Asplund2021}, as reviewed, for instance, in \cite{Catelan2013}, \citet{Bergemann2014}, and \citet{Villante2019}, among many others~-- and indeed, the \cite{Grevesse1998} solar abundances are also more similar to the new values recommended by \citet{2022A&A...661A.140M}.  

The photospheric abundances vary over time, with respect to the protosolar values $Y^{\rm ini}_{\sun}$ and $Z^{\rm ini}_{\sun}$, as a consequence of diffusion, until they reach the current photospheric abundances of the Sun $Y^{\rm surf}_{\sun}$ and $Z^{\rm surf}_{\sun}$. The latter, as we have just seen, are based on \cite{Grevesse1998}. The initial composition is set by assuming a linear enrichment law for the helium abundance, as follows:

\begin{equation}
    Y = Y_p + \left(\frac{Y^{\rm ini}_{\sun}-Y_p}{Z^{\rm ini}_{\sun}}\right)Z,
\label{eq:helium}
\end{equation}
\begin{equation}
    X=1-Y-Z,
\end{equation}

\noindent where $Y_p=0.2437$ is the primordial helium abundance \citep{planck2020}, $Y^{\rm ini}_{\sun}=0.275$, $Z^{\rm ini}_{\sun} = 0.019$, and we adopted $Z=0.005$, $0.007$, and $0.009$ as representative metallicities for LMC stars \citep{Nidever2020}.

\subsection{Microphysics}
\subsubsection{Opacities}
\cite{Ferguson2005} low-temperature tables and OPAL \citep{Iglesias1993,Iglesias1996} high-temperature opacity tables are adopted. The OPAL tables are split into two types. Type I is used for $0.0\le X\le 1.0-Z$ and $0.0\le Z\le 0.1$. Type II tables allow enhanced carbon and oxygen abundances, covering $0.0\le X\le 0.7$ and $0.0\le Z\le 0.1$. Type II opacities are particularly important for helium burning and advanced burning phases.

\subsubsection{Equation of State}
The equation-of-state (EOS) tables in MESA are based on the OPAL EOS tables \citep{Rogers2002}. At lower temperatures and densities, there is a transition to the Saumon-Chabrier-Van Horn (SCVH) tables \citep{Saumon1995}. These extended MESA EOS tables cover $X=0.0,0.2,0.4,0.6,0.8,1$, and $Z=0.0,0.02,0.04$. In addition, for temperatures and densities outside the ranges covered in SCVH, the Helmholtz EOS \citep{Timmes2000} and the Potekhin-Chabrier EOS \citep{Potekhin2010} are used.

\subsubsection{Nuclear Reaction Networks}
We adopted the \texttt{o18\_and\_} \texttt{ne22.net} nuclear network in the version of MESA used throughout this work, using nuclear reaction rates from the Nuclear Astrophysics Compilation of Reaction rates \citep[NACRE,][]{Angulo1999}. This nuclear network tracks and solves for the following species: $^1$H, $^3$He, $^4$He, $^{12}$C, $^{14}$N, $^{16}$O, $^{18}$O, $^{20}$Ne, $^{22}$Ne and $^{24}$Mg. This compact nuclear grid is sufficient for our calculations, which include only the core hydrogen- and helium-burning evolutionary phases.

\subsection{Macrophysics}
\subsubsection{Convection}
Convective energy transport is commonly described by the mixing length theory (MLT), which has a free parameter $\alpha_{\rm MLT}$ that determines how far a fluid parcel travels before dissolving in the medium and depositing its energy. The location of convective regions is determined using the Ledoux criterion. We adopt the version of MLT from \cite{Henyey1965}. This prescription requires two additional free parameters, $\nu$ and $\gamma$, which are multiplicative factors to the mixing length velocity and the temperature gradient in the convective element, respectively. We use a mixing length parameter of $\alpha_{\rm MLT}=1.88$, that was constrained with a solar calibration (Sec.~\ref{subsec:calibration}), and recommended default values of $\gamma=1$ and $\nu=8$ \citep{Paxton2011}. We consider a constant mixing length parameter, without taking into account possible dependencies on, for example, temperature and metallicity \citep{Ludwig1999,Magic2015,Valle2019}.
Convective mixing of elements is treated as a time-dependent diffusive process, with a diffusion coefficient computed within the MLT formalism.

\subsubsection{Convective Overshoot Mixing}
In order to consider the nonzero momentum of a fluid element approaching the edge of the convective zone, as defined by the Ledoux criterion, the convective region is extended beyond the edge thus defined. We adopted the exponential overshooting prescription implemented in MESA, in which the turbulent velocity field decays exponentially out of the convective boundary and eventually the convective elements disintegrate in the overshoot region through a diffusive process. According to the parameterization described in \cite{Herwig2000}, the diffusion coefficient in the extended region is given by:
\begin{equation}
    D_{\rm ov} = D_{\rm 0}\exp\left(-\frac{2z}{f_{\rm ov}H_p}\right),
\end{equation}

\noindent where $f_{\rm ov}$ is a free parameter that determines the efficiency of overshoot mixing, $H_p$ is the local pressure scale height, and $D_{\rm 0}$ is the diffusion coefficient in the convectively unstable region at a depth $f_{0,\rm ov}H_p$ from the convective boundary. We adopt a value of $f_{\rm ov}=0.019$, obtained by performing a solar calibration (Sec.~\ref{subsec:calibration}). For simplicity, we consider that the efficiency of overshooting is the same for the core and the envelope, and $f_{0,\rm ov}$ is set to $0.5 \, f_{\rm ov}$ as in \cite{Choi2016}. In addition, more complex dependencies of this parameter were not considered. For instance,  it has been suggested that the overshooting parameter is a function of mass and metal abundance \citep{Woo2001,VandenBerg2006}, although \cite{Claret2007} showed that the dependence is less pronounced than was suggested by these authors. 

\subsubsection{Semi-convection and Thermohaline Mixing}
Semi-convective mixing occurs in regions that are unstable against convection according to the Schwarzchild criterion but stable according to the Ledoux criterion. The mixing in regions that satisfy the above condition is calculated by a time-dependent diffusive process. The diffusion coefficient is given by the following expression \citep{Langer1983}:
\begin{equation}
    D_{\rm sc} = \alpha_{\rm sc}\left(\frac{K}{6C_{p}\rho}\right)\left(\frac{\nabla-\nabla_{\rm ad}}{\nabla_{\rm L}-\nabla}\right),
\end{equation}
where $K$ is the radiative conductivity, $C_{p}$ is the specific heat at constant pressure, and $\alpha_{\rm sc}$ is a dimensionless efficiency parameter. Semi-convection is important for stars with convective cores, as it can have a significant effect on the latter's size \citep[e.g.,][]{Silva2011,Paxton2013}. Following \cite{Choi2016}, we adopt $\alpha_{\rm sc}=0.1$. 

Thermohaline mixing occurs in the presence of an inversion of the mean molecular weight in regions that are stable against convection according to the Ledoux criterion. In MESA, themohaline mixing is treated in a diffusive approximation, with a diffusion coefficient given by the analysis of \cite{Ulrich1972} and \cite{Kippenhahn1980}:
\begin{equation}
    D_{\rm th}=\alpha_{\rm th}\left(\frac{3K}{2\rho C_p}\right)\left(\frac{B}{\nabla-\nabla_{\rm ad}}\right).
\end{equation}

\noindent The parameter $\alpha_{\rm th}$ is a dimensionless efficiency that depends on the aspect ratio of the blobs or fingers arising from the instability. Themohaline mixing can occur due to the $^3{\rm He}(^3{\rm He},2{\rm p})^4{\rm He}$ reaction, that takes place beyond the H-shell burning region during the red giant branch (RGB), horizontal branch, and asymptotic giant branch (AGB) phases \citep{Eggleton2006,Charbonnel2007}. In the literature, proposed $\alpha_{\rm th}$ values cover two orders of magnitude \citep{Kippenhahn1980,Charbonnel2007,Cantiello2010}. We adopt $\alpha_{\rm th}=667$, since this value reproduces the surface abundance anomalies in RGB stars past the luminosity bump \citep{Charbonnel2007}.

\subsubsection{Boundary Conditions}
The pressure and temperature of the surface layers of a stellar model must be set by boundary conditions. These are set by model atmospheres calculated with the PHOENIX \citep{Hauschildt1999a,Hauschildt1999b} and \cite{Castelli2003} models. These boundary conditions are implemented in MESA in the \texttt{photosphere\_tables} option, which cover $\log (Z/Z_{\odot})=-4$ to $+0.5$, assuming the \cite{Grevesse1993} solar abundance mixture, and they span $\log(g) = -0.5$ to $5.5$ cm s$^{-2}$ and $T_{\rm eff}=2000-50 000$ K. There is a small difference between the abundances adopted to set the boundary conditions \citep{Grevesse1993} and those adopted in our stellar interior calculations \citep{Grevesse1998}. Tests were performed by changing the abundances in the interior of our models to match those used in the boundary conditions. No significant differences in the resulting evolutionary tracks and pulsation properties of the models were observed.

\subsubsection{Diffusion}
Diffusion and gravitational settling of elements are essential in models of stellar evolution, causing modifications to the surface abundances and duration of the MS phase, as well as a shift in the evolutionary tracks to low luminosities and temperatures in the HRD, when diffusion is considered \citep{Michaud1984,Salaris2000,Chaboyer2001,Stancliffe2016}. MESA performs diffusion and gravitational settlement calculations following the method of \cite{Thoul1994}. The elements present in the stellar model are categorized into five ``classes'' according to their atomic mass, each of which has a representative element whose properties are used to calculate the diffusion velocities. We adopt the default MESA set for representative members; these are $^1$H, $^3$He, $^4$He, $^{16}$O, and $^{56}$Fe. Atomic diffusion coefficients are calculated following \cite{Paquette1986}. Then, the diffusion equation is solved using the total mass fraction within each class.

\subsubsection{Rotation}\label{subsec:rotation}
The rotation of stars has been widely studied from an evolutionary perspective \citep[e.g.,][]{Pinsonneault1990,Maeder2000,Heger2000,Ekstrom2012,Georgy2013,Eggenberger2021}, but its effects on models of stellar evolution remain an uncertain problem. Stellar structure deviates from spherical symmetry in the presence of rotation. While the structure is inherently 3D, it is possible to solve the stellar structure equations in one dimension by assuming the ``shellular approximation'' \citep{Kippenhahn1970,Meynet1997,Paxton2013}. This approach is valid if the angular velocity is constant over isobars, which is to be expected in the presence of strong anisotropic turbulence acting on these isobars due to differential rotation \citep{Zahn1992}.

In MESA, modifications to the stellar structure equations are made by introducing two correction factors, namely $f_P$ and $f_T$, to the momentum balance and the radiative temperature gradient \citep[see][]{Paxton2013,Paxton2019}. Previous versions of MESA used the \cite{Endal1976} method, which considers deviations of the Roche potential from spherical symmetry, to calculate $f_P$ and $f_T$. These parameters need a minimum value to ensure numerical stability ($f_P=0.75$ and $f_T=0.95$). \cite{Paxton2019} implemented analytical fits to the Roche potential, that do not need to establish a minimum for $f_P$ and $f_T$, in order to improve the calculations of models with high rotation. However, after numerous tests carried out in this work, numerical stability was not achieved using these analytical fits for models with $\omega_0 \equiv\Omega_{\rm ZAMS}/\Omega_{\rm crit}>0.5$, where $\Omega_{\rm crit}$ is the critical angular frequency, defined as
\begin{equation}
    \Omega_{\rm crit}=\sqrt{GM/R_{\rm eq}^3},
\end{equation}

\noindent and $R_{\rm eq}$ is the equatorial radius when the star reaches critical rotation. Therefore, minimum values of $f_p=0.75$ and $f_T=0.95$ were adopted, which correspond to a maximum rotation rate of $60$ per cent of the critical rotation. Results for high-rotation models should accordingly be treated with caution, since corrections to the structure equations may be underestimating the effects of rotation.

Initial rotation is defined on the zero-age main sequence (ZAMS) as solid-body rotation. The input parameter for varying the rotation rate is the ratio of the initial surface angular frequency to the critical angular frequency $\omega_0$. In this work, we calculate stellar evolution models with $\omega_0=0.1$, $0.2$, $0.3$, $0.4$, $0.5$, $0.6$, $0.7$, $0.8$, $0.9$. We stress that the results for $\omega_0>0.6$ should be treated with caution, due to the minimum values adopted for correction parameters due to rotation.

The transport of angular momentum and chemical elements due to rotation-induced instabilities is implemented in MESA in a diffusive approach, as described in \cite{Endal1978}. It is important to note that other codes of stellar evolution, such as the Geneva  \citep{Eggenberger2008} and RoSE \citep{Potter2012} codes, implement a diffusion-advective description \citep{Zahn1992} that is different from MESA's. These two approaches are equivalent for the transport of elements, but can cause great differences in the transport of angular momentum. MESA calculates diffusion coefficients for five rotationally-induced mixing processes: dynamical shear instability, Solberg-H\o iland instability, secular shear instability, Eddington-Sweet circulation, and the Goldreich-Schubert-Fricke instability. A detailed description of the physics of these phenomena and the calculation of the respective diffusion coefficients can be found in \cite{Heger2000(2)}. This diffusive implementation of the transport of angular momentum and chemical elements has two free parameters, namely $f_c$, which represents the ratio of the diffusion coefficient to the turbulent viscosity, and scales the efficiency of composition mixing to that of angular momentum transport, and $f_\mu$, which relates the sensitivity of the rotational mixing to the mean molecular weight gradient. In other words, a small $f_c$ corresponds to a process that transports angular momentum more efficiently than it can mix material, and a small $f_\mu$ means that the rotational mixing is efficient even in the presence of a stabilizing gradient of the molecular weights in the star. We consider $f_c = 1/30$ and $f_{\mu}=0.05$ after \cite{Choi2016}, who demonstrate that these values produce surface nitrogen enhancements that are in reasonable agreement with observations.

\subsubsection{Mass Loss}\label{sec:massloss}
RGB mass loss is treated with the \cite{Reimers1975} prescription. This scheme depends on global stellar properties as follows:

\begin{equation}
    \dot{M}_{\rm R}=4\times10^{-13}\eta_{R}\frac{(L/L_{\odot})(R/R_{\odot})}{(M/M_{\odot})}{\rm M}_{\odot}{\rm yr}^{-1},
\end{equation}

\noindent where $\eta_{R}$ is a scaling factor that represents the mass loss efficiency. As in \cite{Anderson2016}, we adopt a value of $\eta_{\rm R} = 0.5$ for stars within the range of $4\le M/{\rm M}_{\odot}\le 5$, and $\eta_{\rm R}=0.6$ for stars within the range of $5.5\le M/{\rm M}_{\odot}\le 7$. Since our models do evolve up the AGB, for completeness we state also that we consider mass loss in the AGB phase as given by the scheme of \cite{Bloecker1995},

\begin{equation}
    \dot{M}_{\rm B}=4.83\times10^{-9}\eta_{\rm B}\frac{(L/L_{\odot})^{2.7}}{(M/M_{\odot})^{2.1}}\frac{\dot{M}_{{\rm R}}}{\eta_{{\rm R}}}{\rm M}_{\odot}{\rm yr}^{-1},
\end{equation}

\noindent where $\eta_{\rm B}=0.0003$. Since our interest lies in the stages prior to the AGB, this assumption does not affect any of our results. 

In addition to the above prescriptions, MESA includes rotationally-enhanced mass loss, expressed as a function of the surface angular frequency $\Omega$ as follows:

\begin{equation}
    \dot{M}(\Omega) = \dot{M}(0)\left(\frac{1}{1-\Omega/\Omega_{\rm crit}}\right)^{\xi},
\end{equation}

\noindent where $\dot{M}(0)$ is the standard mass-loss rate (Reimers or Blöcker), and we adopt $\xi=0.43$ after \citet{Langer1998}. Some of our models reached critical rotation during their evolution; in such cases, we expect a strong increase in mass loss in the equatorial region. The exact details of this process are still uncertain and require simulations that combine hydrodynamics and radiative transfer. In response to this situation, MESA implements a ``mechanical mass loss,'' which removes the super-critical layers and ensures that the surface is kept below critical velocity.

\subsection{Solar Calibration}\label{subsec:calibration}
\begin{table}
    \centering
    \begin{tabular}{c|c|c|c}
        \hline
        Parameter&Target&Model value&Fractional error (\%)\\\hline
        $L_{\odot}$ ($10^{33}$ erg s$^{-1}$)& 3.828$^{a}$& 3.823&0.13 \\
        $R_{\odot}$ ($10^{10}$ cm)&6.957$^{a}$& 6.957&0.005\\
        $T_{\rm eff,\odot} $ (K)&5772$^{a}$& 5774 &0.03 \\
        $Y^{\rm surf}_{\odot}$ &0.2485$^{b}$ & 0.2503 &0.72\\
        $Z^{\rm surf}_{\odot}/X^{\rm surf}_{\odot}$ &0.0231$^{c}$ &0.0248 &7.36\\
        $R_{\rm cz}$ &0.713$^{b}$ & 0.717 & 0.56\\\hline
        $\alpha_{\rm MLT}$ & ... & 1.88 & ...\\
        $f_{\rm ov}$ & ... & 0.019 & ...\\\hline
    \end{tabular}
    \caption{Solar calibration results. $^{a}$ \protect\cite{Mamajek2015}, $^{b}$ \protect\cite{Basu2004}, $^{c}$ \protect\cite{Grevesse1998}.}
    \label{tab:table2_1}
\end{table}

As mentioned in the previous sections, a calibration of the mixing length and overshooting parameters was performed using the constraints provided by the location of the base of the Sun's convective zone $R_{\rm cz}$, as obtained using helioseismic data \citep{Basu2004}, and surface properties of the Sun \citep{Mamajek2015}. We used the \texttt{simplex\_solar\_calibration} test suite, which uses the simplex optimization algorithm \citep{Nelder1965}, to find models that minimize a specific $\chi^2$ to obtain a set of parameters that reproduce observationally inferred solar parameters. For each iteration, a new choice of the mixing length and overshooting parameters is set, and the model is evolved from the pre-MS up to an age of 4.61~Gyr. A $\chi^2$ value is computed by summing over the residuals between model and observational $\log L$, $\log R$, $T_{\rm eff}$, surface abundance, and $R_{\rm cz}$ values. This process is repeated until the tolerance parameters are met.

The results of the solar calibration are shown in Table~\ref{tab:table2_1}. Although many initial assumptions and variations of diffusive parameters were explored, we did not obtain a model that satisfies all available solar observations \citep[see also, e.g.,][]{Catelan2013,Serenelli2016}. In particular, the largest discrepancies are found in the ratio of surface abundances $Z^{\rm surf}_{\odot}/X^{\rm surf}_{\odot}$.

We accordingly adopt a solar-calibrated $\alpha_{\rm MLT}=1.88$ for all masses and an overshoot parameter for the envelope and the core $f_{\rm ov, env}=f_{\rm ov,core}=0.019$ ($f_{0,\rm ov, env}=f_{0,\rm ov, core}=0.5f_{\rm ov, env}$).

\section{Stellar Evolution Models}\label{sec:Models}
We calculate evolutionary tracks that cover stellar masses from $4$ to $7 \, {\rm M}_{\odot}$, uniformly spaced in mass with steps of $0.5\,{\rm M}_\odot$, metallicities of $Z=0.005$, $0.007$, and $0.009$, helium abundances computed with equation~\eqref{eq:helium} of $Y=0.252$, $0.256$ $0.259$, respectively, and a rotation rate $\omega_0$ ranging from 0.0 to 0.9. The evolution is calculated from the pre-MS stage to a limit in luminosity that allows the thermal-pulsing AGB phase to be avoided. The latter limit was implemented since post-AGB phases are computationally expensive and are not relevant to our study of classical Cepheids.

\begin{figure*}
    \centering
    \includegraphics[scale=0.64]{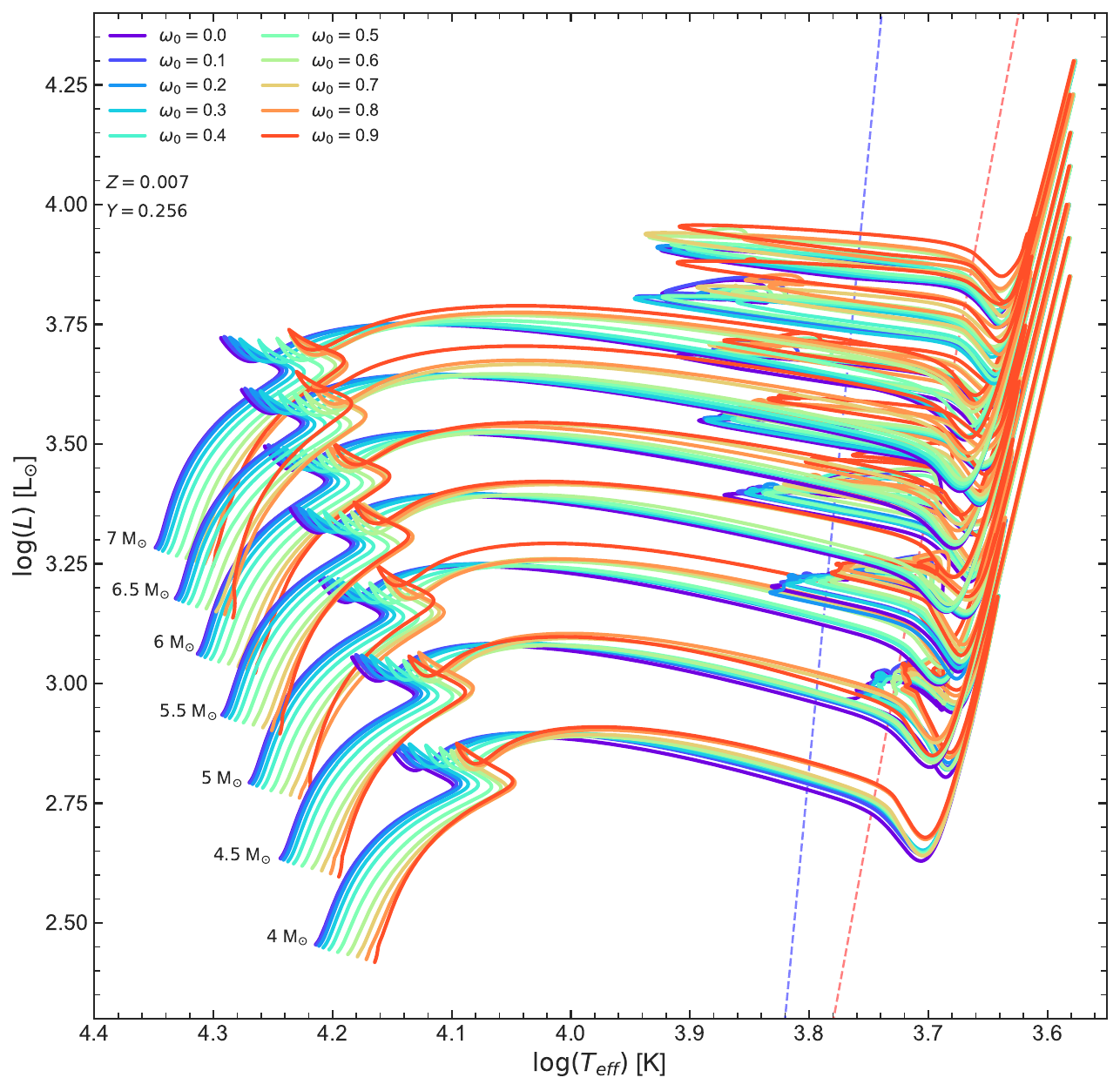}
    \caption{HRD of the evolutionary tracks computed with MESA, assuming a metallicity $Z=0.007$ and a helium abundance $Y=0.256$. Lines of different colors correspond to different $\omega_0$, following the color scheme given at the top left. Averaged IS edges for fundamental-mode radial pulsation, as calculated with RSP, are plotted as dashed lines. Pre-MS evolution is omitted for clarity.}
    \label{fig:fig3_1}
\end{figure*}

\begin{figure*}
    \centering
    \includegraphics[width=\hsize]{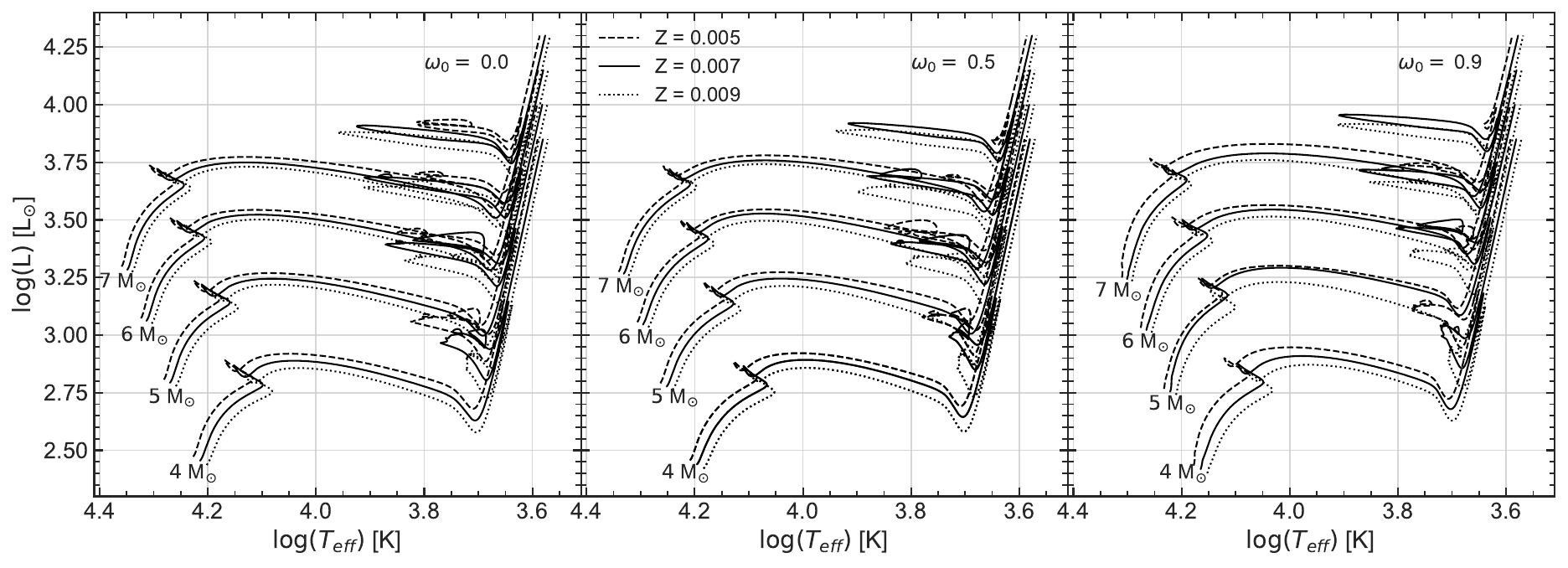}
    \caption{HRD showing the computed evolutionary tracks, assuming $\omega_0 = 0.0$ (left panel), $0.5$ (middle), and $0.9$ (right). In each panel, tracks are shown for masses, from bottom to top, of $4$, $5$, $6$, and $7 \, {\rm M}_{\odot}$, respectively. Results for three different metallicities, namely $Z = 0.005$ (dashed lines), $0.007$ (solid lines), and $0.009$ (dotted lines)  are also displayed in each panel.}
    \label{fig:fig3_1_2}
\end{figure*}

\subsection{Basic Properties of the Models}\label{sec:properties}
\subsubsection{HR Diagram}\label{sec:HDR}
\begin{figure}
    \centering
    \includegraphics[width=\hsize]{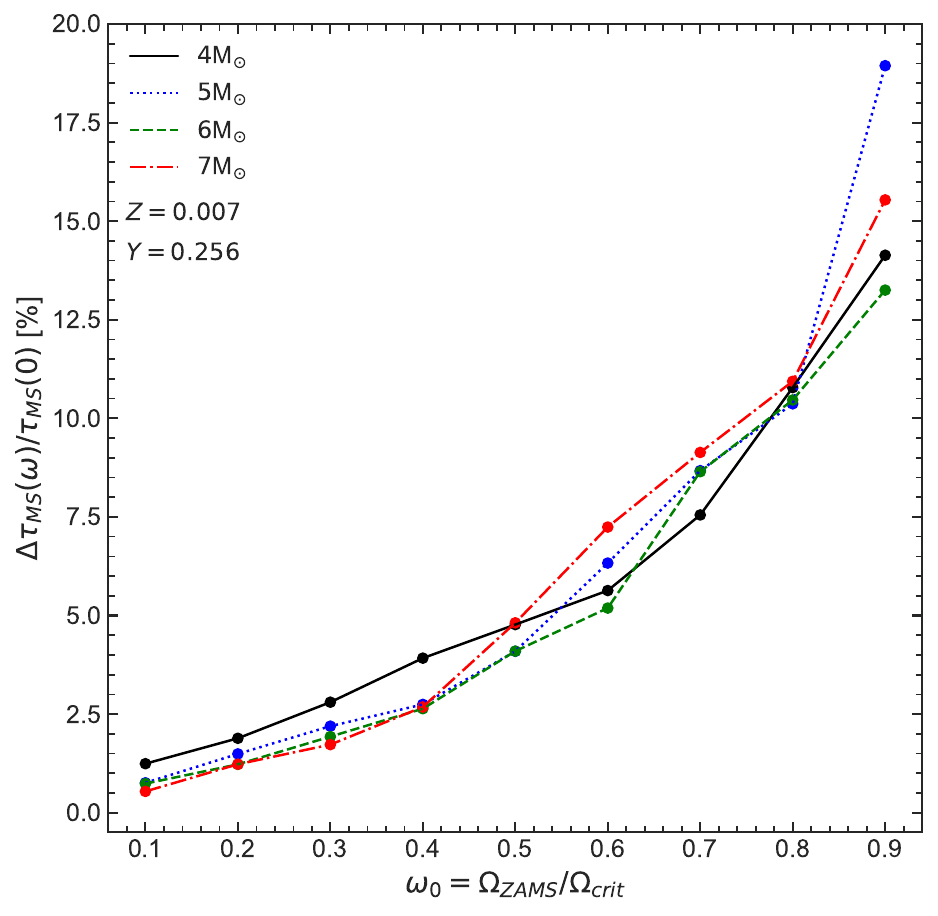}
    \caption{MS duration enhancement, computed for models with metallicity $Z=0.007$ and helium abundance $Y=0.256$, as a function of the initial rotation rate parameter $\omega_0$. Lines of different colors and styles correspond to different masses, namely $4$ (black solid line), $5$ (blue dotted line), $6$ (green dashed line), and $7 \, {\rm M}_{\odot}$ (red dashed-dotted line).}
    \label{fig:fig3_3}
\end{figure}

Figure~\ref{fig:fig3_1} shows the calculated evolutionary tracks, for $Z=0.007$. Selected tracks with metallicities $Z=0.005$ and $Z=0.009$ are shown in Figure~\ref{fig:fig3_1_2}. For each of the considered masses, the temperature of the ZAMS becomes progressively lower as the rotation rate increases, with differences  of up to $\sim1700$ K. In rotating models, this initial evolutionary phase is dominated by the centrifugal force \citep{Meynet2000}, which decreases the effective gravity acceleration $g_{\rm eff}$. Since $T_{\rm eff}\propto g_{\rm eff}^{1/4}$, in a rotating star at the ZAMS the effective temperature is expected to be lower \citep{Zeipel1924}. During the MS, the behavior of the luminosity as a function of rotation is related to the mass of the convective core, which is affected by a competition between two physical effects. On one hand, the centrifugal force generates additional support to gravity, which tends to decrease the size of the core and its luminosity. On the other hand, the opposite occurs due to rotational mixing, which brings hydrogen-rich material to the convective core, slowing down its decrease in mass and extending the duration of the MS, as shown in Figure~\ref{fig:fig3_3}, leading to an extension of its duration by $\sim15$ per cent for a $\omega_{0}=0.9$. In addition, rotational mixing transports helium and nitrogen to the radiative envelope. This material decreases the opacity of that medium, which produces an increase in the luminosity.

At the same stellar age, a non-rotating model is brighter than a rotating one, indicating that the centrifugal force is more effective at decreasing the core size than is rotational mixing at increasing it. The behavior of the luminosity is opposite to the one previously mentioned during the H-shell burning phase, since we notice that, as the initial rotation rate increases, the evolutionary tracks become more luminous. 

\begin{figure*}
    \centering
    \includegraphics[scale=0.54]{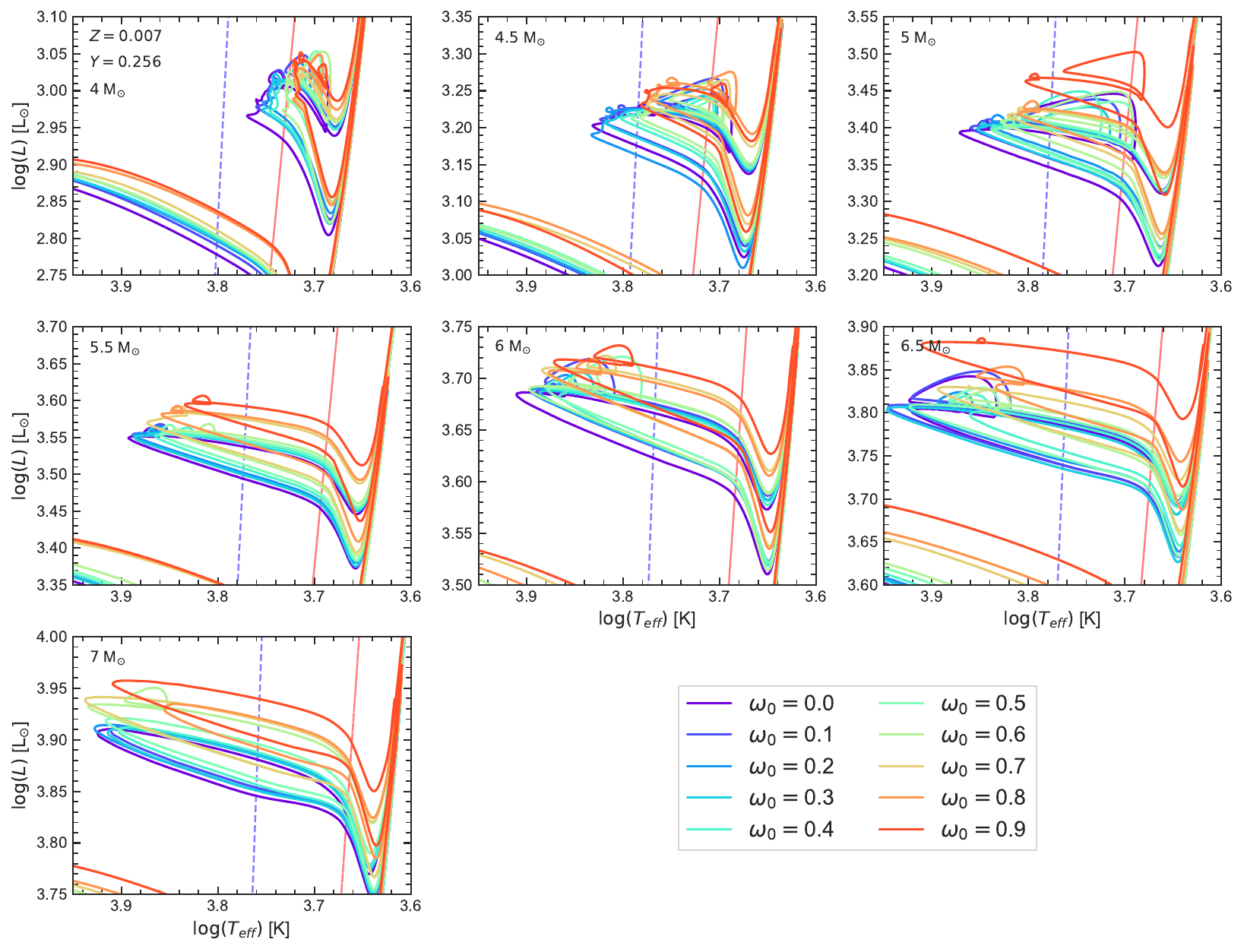}
    \caption{Evolution in the HRD of models with metallicity $Z=0.007$ and helium abundance $Y=0.256$ during core-He burning. Red and blue averaged IS boundaries for fundamental-mode pulsation, computed using RSP, are overplotted. From upper left to bottom left, results are showm for masses from $4$ to $7 \, {\rm M}_{\odot}$ in steps of $0.5 \, {\rm M}_{\odot}$, with mass values increasing to the right along each row. Lines of different colors correspond to different $\omega_0$, following the color scheme given at the bottom right of the figure.}
    \label{fig:fig3_2}
\end{figure*}

A closer view of the evolution of the models during the blue loop is shown in Figure~\ref{fig:fig3_2}. These loops are the consequence of an excess of helium above the H-burning shell, which results from the contraction of the convective core during core hydrogen burning. The outward movement of the burning shell removes the excess helium and produces the loop \citep{Walmswell2015}. Blue loops are very sensitive to metallicity and the adopted input physics, such as convective core overshooting and nuclear reactions \citep{2004Xua,2004Xub,Walmswell2015}. In general, we notice that there is an increase in the luminosity of the loops as the rotation rate increases. For $M\le 5.5 \, {\rm M}_{\odot}$, the blue loop extension decreases as the rotation rate increases. However, this does not occur in tracks with $M> 5.5 \, {\rm M}_{\odot}$, where the loop extension behavior is non-monotonic. A similar phenomenon can be observed in the models presented in \cite{Georgy2013}. As a consequence, low-mass, high-rotation Cepheids are less likely to be observed since they spend less time during the blue loop phase.

We note that some tracks, especially the $4-5$ ${\rm M}_{\odot}$ ones, exhibit multiple smaller loops during the He-burning phase. These are commonly called ``He-spikes,'' which in some cases produce a fourth, and even a fifth, crossing of the IS. The presence of these smaller loops is directly associated with sudden increases in the central helium abundance. It is unlikely that these He-spikes are real, although small mixing events within a semiconvective zone at the edge of the convective core, which move a small amount of helium inwards \citep{Sweigart1979}, could produce a similar effect. In MESA, these episodes are associated with the uncertainty in the location of the edge of the convective core, and the mixing that occurs at this interface. A small displacement of this boundary can mix a significant amount of helium in the core, which produces an increase in energy generation. This ends up producing noticeable changes in the luminosity of the star. As far as the present study is concerned, the main effect of these He-spikes is the production of a fourth and a fifth crossing of the IS. In addition, it has also been noted that He-spikes produce alterations of the lifetimes for the third crossing, which can produce discrepancies in the predictions for the number ratio of AGB to HB stars in globular clusters \citep{Constantino2017}. 

The version of MESA used in this work implements a new approach for the treatment of convective boundaries, called the ``convective premixing (CPM)'' scheme \citep{Paxton2019}. The CPM scheme is applied at the start of each time step, before any structural or compositional changes in the previous step. It finds the boundary cells between the convective and radiative zones, and considers whether the radiative face of the cell would change if the adjacent cell outside the convective region were completely mixed with the rest of the convective region. If the radiative face of the boundary cell becomes convective during this putative mixing, the mixing is applied in the model. This process continues iteratively, until the radiative face of the current convective boundary remains radiative during the putative mixing.

As can be seen in Figure~\ref{fig:fig3_1_2}, small changes in $Z$ can lead to significant changes in the positions of the tracks in an HRD. During core- and shell-hydrogen burning, the decrease in metal content makes the evolutionary track hotter and more luminous, due to a lower Rosseland mean opacity. During core-He burning, differences in stellar envelope opacity play an important role in the behavior of the blue loops. \cite{Walmswell2015} concluded that, as a rule, the higher the metallicity, the less pronounced the blue loop becomes. However, the sensitivity of the loop extension to the input physics adds a higher degree of complexity. In Figure~\ref{fig:fig3_1_2}, we observe how this general trend holds for stars of $4 \, {\rm M}_{\odot}$; however, the behavior for more massive stars is more intricate, with the blue loop being completely suppressed for stars with $7  \, {\rm M}_{\odot}$, metallicity of $Z=0.005$, and initial rotation rates of $\omega_{0}=0.5$ and $0.9$. There is a number of important ingredients that determine whether a star develops a blue loop or not. Among those, the ratio of the mass of the convection zones within the envelope to the total envelope mass at
the bottom of the RGB is essential. Stellar models with very low or high metallicity are found to develop blue loops when this ratio is lower than a critical value between $0.3$ and $0.45$ \citep{2004Xua,2004Xub}. Clearly, the development, extension, and detailed morphology of the blue loop constitute a complex problem, whose detailed analysis is beyond the scope of this paper.

\subsection{Evolution of the Surface Rotational Frequency}\label{sec:omega}
\begin{figure*}
    \centering
    \includegraphics[scale=0.48]{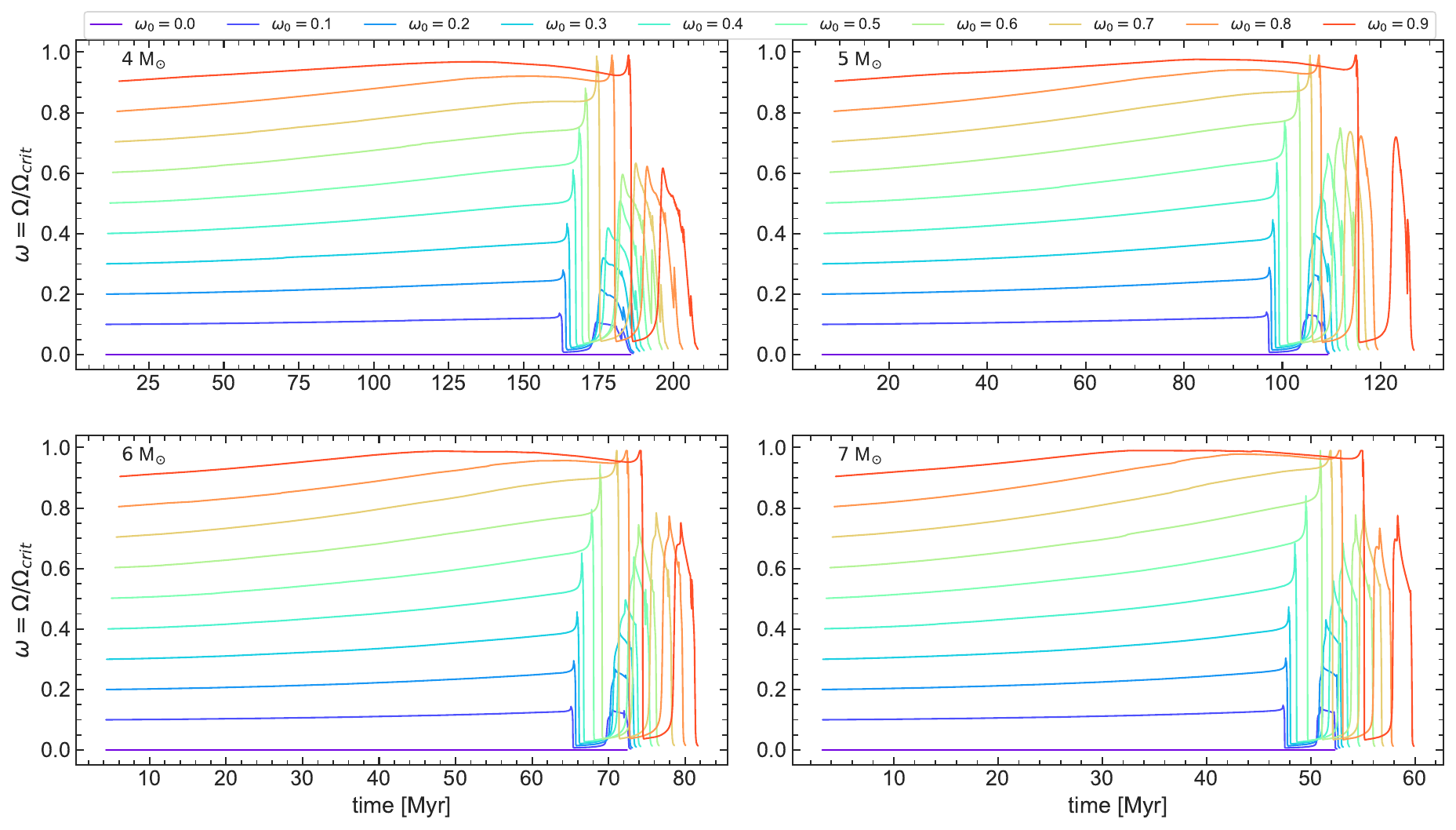}
    \caption{Evolution of the surface angular frequency of rotation for $4$ (upper left), $5$ (upper right), $6$ (bottom left), and $7 \, {\rm M}_{\odot}$ (bottom right). In each panel, lines of different colors correspond to different $\omega_0$, following the color scheme given at the top of this figure.}
    \label{fig:fig3_5}
\end{figure*}

The evolution of the surface rotational frequency is shown in Figure~\ref{fig:fig3_5}. Three physical processes are responsible for the evolution of the rotation rate: conservation of angular momentum, which modifies $\omega$ when the star contracts or expands; internal transport mechanisms \citep{Endal1978}, which redistribute the angular momentum along the stellar interior; and mass loss, which removes angular momentum from the surface.

During the MS, Eddington circulation is one of the most important mechanisms changing the angular momentum distribution \citep{Endal1978}, as it carries angular momentum from the inner parts to the surface, accelerating the latter. On the other hand, mass loss increased by rotation removes angular momentum \citep{Langer1998}. In Figure~\ref{fig:fig3_5}, we notice how $\omega$ stably increases during the MS, with the exception of the tracks with $\omega_0\ge0.8$, for which a small decrease in $\omega$ is observed at the end of the MS. At these high rotation rates, the increased mass loss due to rotation becomes considerable, decreasing the rotation rate by removing angular momentum.

After central hydrogen exhaustion, an overall contraction of the star occurs, which considerably increases $\omega$. For models with high rotation rates, this contraction increases the surface angular frequency up to the critical frequency. When this occurs, the mechanical mass loss mechanism (see Sect.~\ref{sec:massloss}) is activated, keeping the surface below the critical frequency limit by removing the outermost layers. During the H-shell burning phase, the rapid expansion of the star envelope dramatically decreases $\omega$. During He-burning, $\omega$ increases again until the bluest part of the blue loop, then decreases as the star evolves towards the AGB. We note that, for $\omega_{0}\leq0.7$, $\omega$ becomes comparable to or even larger than $\omega_{0}$ during helium burning. On the other hand, for $\omega_0>0.7$, the maximum rotational angular frequency at the surface is always smaller than $\omega_{0}$, and remains constant for all masses. This is due to conservation of angular momentum during this evolutionary phase. A slight increase in $\omega$  is observed in stars with He-spikes, similar to the behavior of the convective core mass. It follows from Figure~\ref{fig:fig3_5} that it would not be possible to observe a Cepheid with a near-critical rotation rate in this mass range due to the loss of angular momentum during the H-shell burning phase, in addition to the enhanced mass loss by rotation during the helium-burning phase.

\subsection{Comparison with Anderson et al. (2016)}\label{subsec:anderson}
\cite{Anderson2016} present the first detailed pulsational instability analysis of stellar evolution models that include rotation, for fundamental mode and first overtone classical Cepheids alike. They use the Geneva code of stellar evolution \citep{Eggenberger2008}, varying mass, metallicity, and rate of rotation. Since the pulsation analysis presented in \cite{Anderson2016} was performed for metallicities $Z=0.014$, $0.006$, and $0.002$, we used the interpolation interface provided by the Synthetic Clusters Isochrones $\&$ Stellar Tracks (SYCLIST) web portal \footnote{\url{https://www.unige.ch/sciences/astro/evolution/en/database/syclist/}}. This web resource performs an interpolation on the grid of stellar models presented in \cite{Georgy2013}, and was used by us to obtain evolutionary tracks for stars of masses $5$ and $7 \, {\rm M}_{\odot}$, metallicity $Z=0.007$, and rotation rates $\omega_0 = 0.0$, $0.5$, and $0.9$.

\begin{figure*}
    \centering
    \includegraphics[width=\hsize]{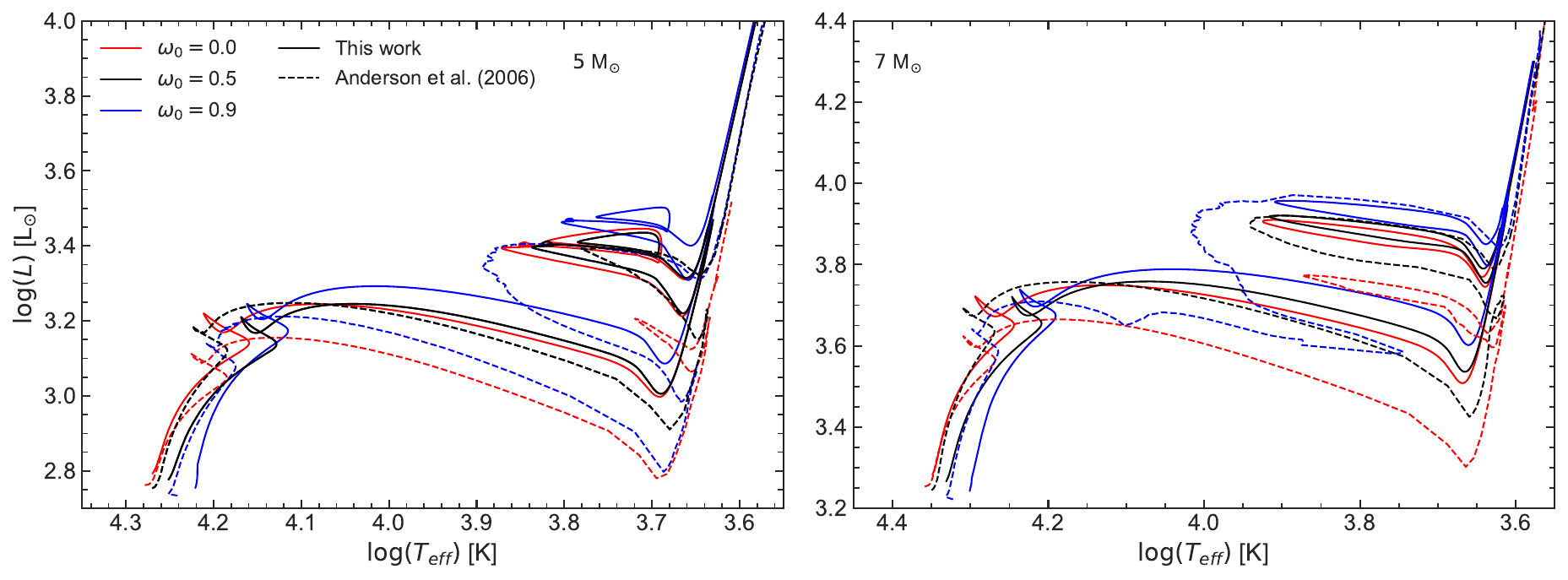}
    \caption{HRD comparing evolutionary tracks of this work (solid lines) with tracks from \protect\cite{Anderson2016} (dashed line). Left panel shows $5 \, {\rm M}_{\odot}$ tracks, while right panel shows $7 \, {\rm M}_{\odot}$ tracks. In each panel, lines of different colors correspond to different $\omega_0$, namely $\omega_0 = 0$ (red), $\omega_0 = 0.5$ (black) and $\omega_0 = 0.9$ (blue).}
    \label{fig:fig3_8}
\end{figure*}

A plot comparing the evolutionary tracks of $5$ and $7 \, {\rm M}_{\odot}$ obtained in this work and in \citet{Anderson2016} in an HRD is shown in Figure~\ref{fig:fig3_8}. We notice important differences between both sets of models, such as in the position of the ZAMS, the extension of the MS, the H-shell burning phase, the position of the RGB, and the behavior of the blue loop. \cite{Anderson2016} uses \cite{Georgy2013} models, which adopted a mixing length of $\alpha_{\rm MLT}=1.65$ and an implementation of instantaneous overshooting with a parameter of $f_{\rm ov}=0.1$. These parameters were calibrated to the Sun, but no details of the calibration process are given in their work. In addition, the temporal resolution of the models calculated in this work is much higher than the resolution of the models of \cite{Anderson2016}: the time step of the models calculated with MESA is between $1$ to $4$ orders of magnitude lower than the time step used to compute the \cite{Anderson2016} models.

\begin{figure}
    \centering
    \includegraphics[width=\hsize]{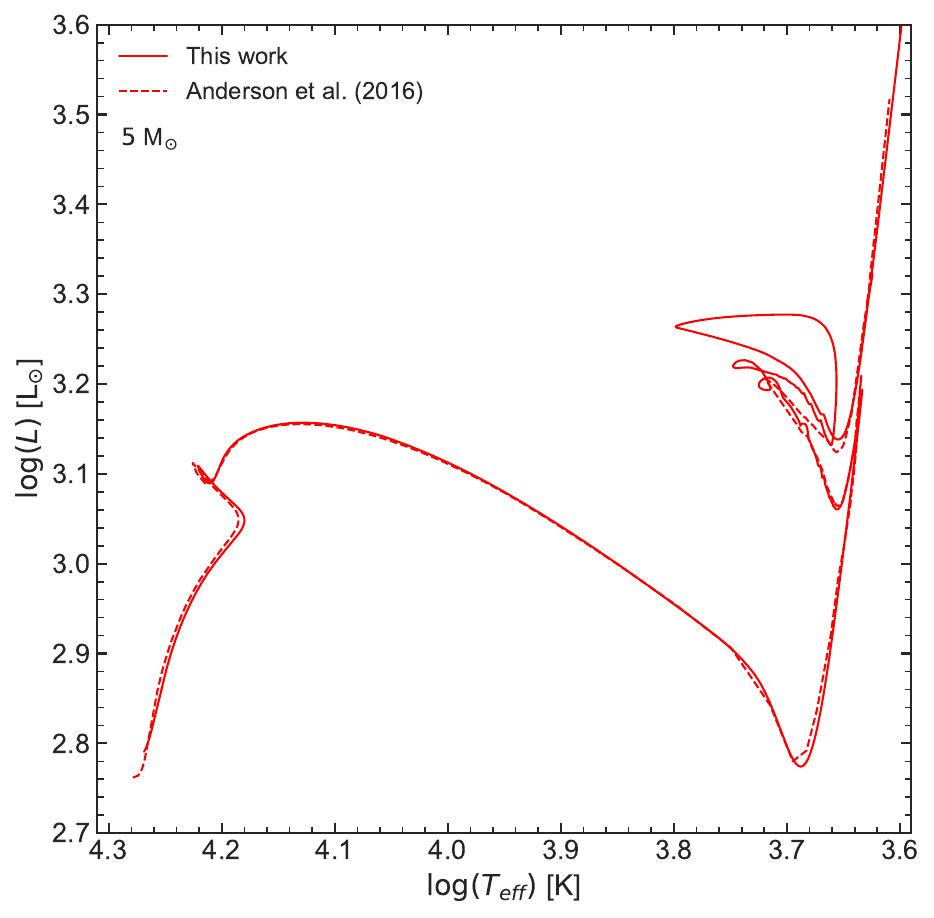}
    \caption{HR diagram comparing a non-rotating track of $5 \, {\rm M}_{\odot}$ from this work (solid lines) with a track from \protect\cite{Anderson2016} with the same mass and chemical composition (dashed line). The same parameters were adopted for the mixing length and instantaneous overshooting, leading to a close match to Anderson's track.}
    \label{fig:fig3_9}
\end{figure}

We created a non-rotating evolutionary model with MESA adopting the same mixing length and overshooting parameters as in \cite{Anderson2016}. An HRD comparing our track with theirs is shown in Figure~\ref{fig:fig3_9}. We notice that the tracks are very similar, with the exception of the behavior in the blue loop, where the track calculated with MESA presents a He-spike, due to the different treatment of the convective boundary. However, matching these parameters was not enough to eliminate the differences between our rotating models and those from \citet{Anderson2016}. This is related to the different implementations of rotation in the Geneva and MESA codes, as mentioned in Section~\ref{subsec:rotation}.

\begin{figure}
    \centering
    \includegraphics[width=\hsize]{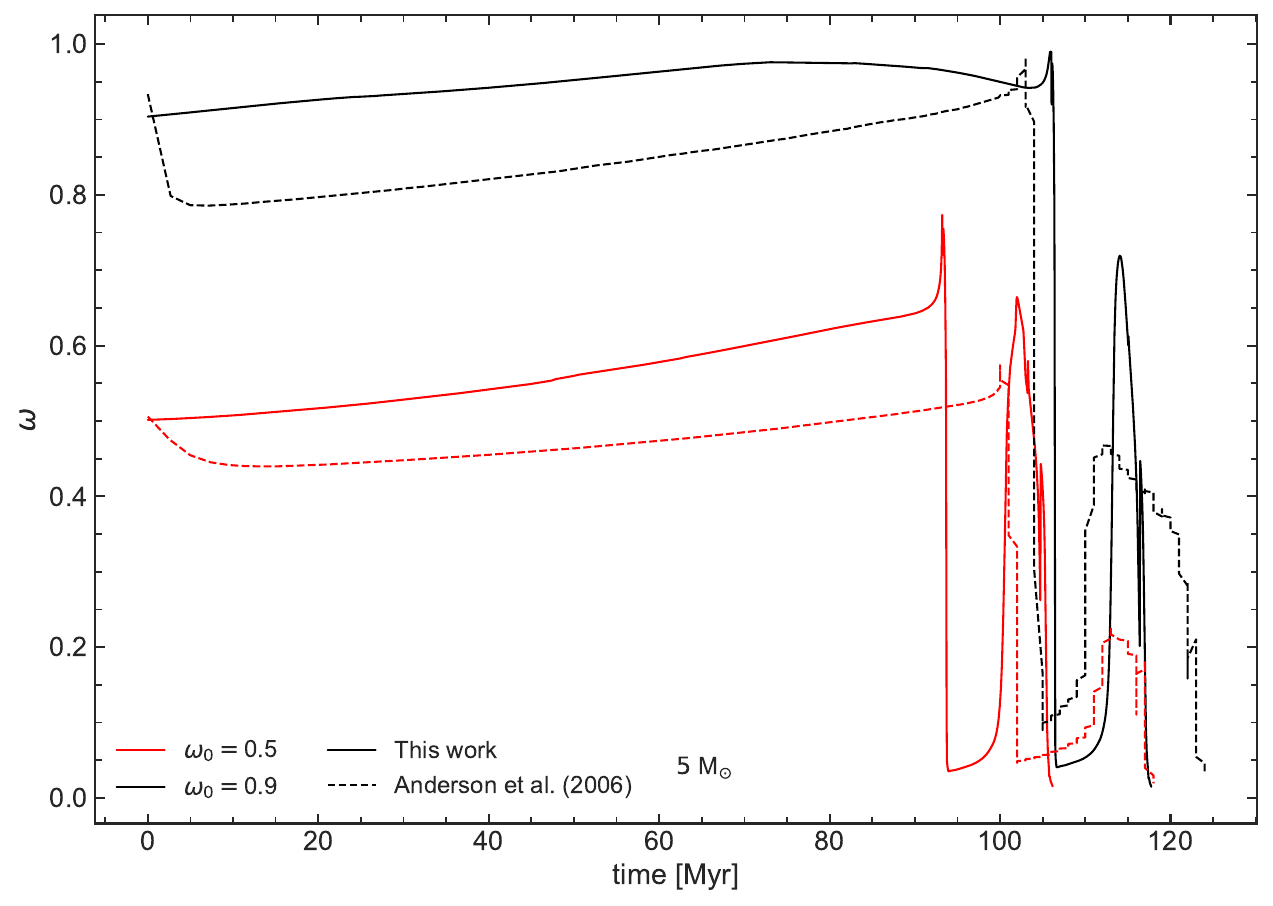}
    \caption{Comparison of the evolution of the angular frequency of rotation between models with $5\,{\rm M}_{\odot}$ computed in this work (solid lines) and \protect\cite{Anderson2016} models of similar composition and mass (dashed line), for two rotation rates.}
    \label{fig:fig3_10}
\end{figure}

The impact of these different implementations can be seen in Figure~\ref{fig:fig3_10}, where the evolution of the rotation rate is shown. A sudden initial decrease in $\omega$ is found in the  \cite{Anderson2016} models. This corresponds to the time that it takes their model to relax towards a state of quasi-equilibrium, from an initially assumed solid-body configuration. In our models, this does not occur since the rotation rate is set near the ZAMS, and the model relaxes over a number of steps until the desired rate is reached. The evolution of $\omega$ during the helium-burning phase also differs between our models and theirs. We note that $\omega$ in the bluest point of the blue loop is higher in our models, thus our evolutionary tracks produce helium-burning stars with higher rotation rates than those predicted by \cite{Anderson2016}. These differences can result in discrepancies in the calculated PCR values, as will be discussed in Section~\ref{sec:comparison}.


\section{Period Change Rates using MESA}\label{sec:pdotMESA}
This section describes how PCRs were obtained using the RSP functionality of MESA. These rates were calculated for each evolutionary track described in the previous section. We study the effects of an increase in the rotation rate, and the calculated PCR values are compared with those recently measured by \cite{Rodriguez2021} for LMC Classical Cepheids.

\subsection{Radial Stellar Pulsation}
\texttt{RSP} is a recent functionality added to MESA \citep[][ adopting the pulsational code of \citealt{Smolec2008}]{Paxton2019} that models high-amplitude, self-excited, non-linear pulsations that the star develops when it crosses the IS. \texttt{RSP} performs three operations: it generates an initial model of the envelope, performs a linear non-adiabatic stability analysis on the model, and integrates the time-dependent non-linear equations. As a result, we obtain a model of the non-linear radial pulsations, growth rates of the three lowest-order radial pulsation modes, and linear periods of the excited pulsation modes.

Since the inner parts of the star do not participate in the oscillations of classical pulsators, a complete stellar model is not necessary \citep{Smolec2008,Paxton2019}. Therefore, \texttt{RSP} is currently limited to pulsations determined by the structure of the stars' envelopes. In our work, the latter are based on the appropriate combinations of $M$, $L$, $T_{\rm eff}$, $X$, and $Z$, guided by approximate IS edges (as provided in MESA's test suite \texttt{rsp\_check\_2nd\_crossing}) and the relevant parameters of our stellar models, as they cross the IS. The \texttt{RSP} model depends on equations describing time-dependent convection, described in \cite{Smolec2008}, which also depend on free parameters that are listed in Table 3 of \cite{Paxton2019}. Pulsation periods depend weakly on these parameters. However, period growth rates and light curves are sensitive to the choice of these convective variables. Different sets for these parameters are shown in Table 4 of \cite{Paxton2019}. Set A corresponds to the simplest convective model, set B adds radiative cooling, while set C adds turbulent pressure and turbulent flux. Set D, which includes these same effects in addition to radiative cooling, showed convergence problems and was therefore discarded. A comparison of convective parameter sets is made in Section~\ref{subsec:parameters}

\begin{figure}
    \centering
    \includegraphics[scale=0.5]{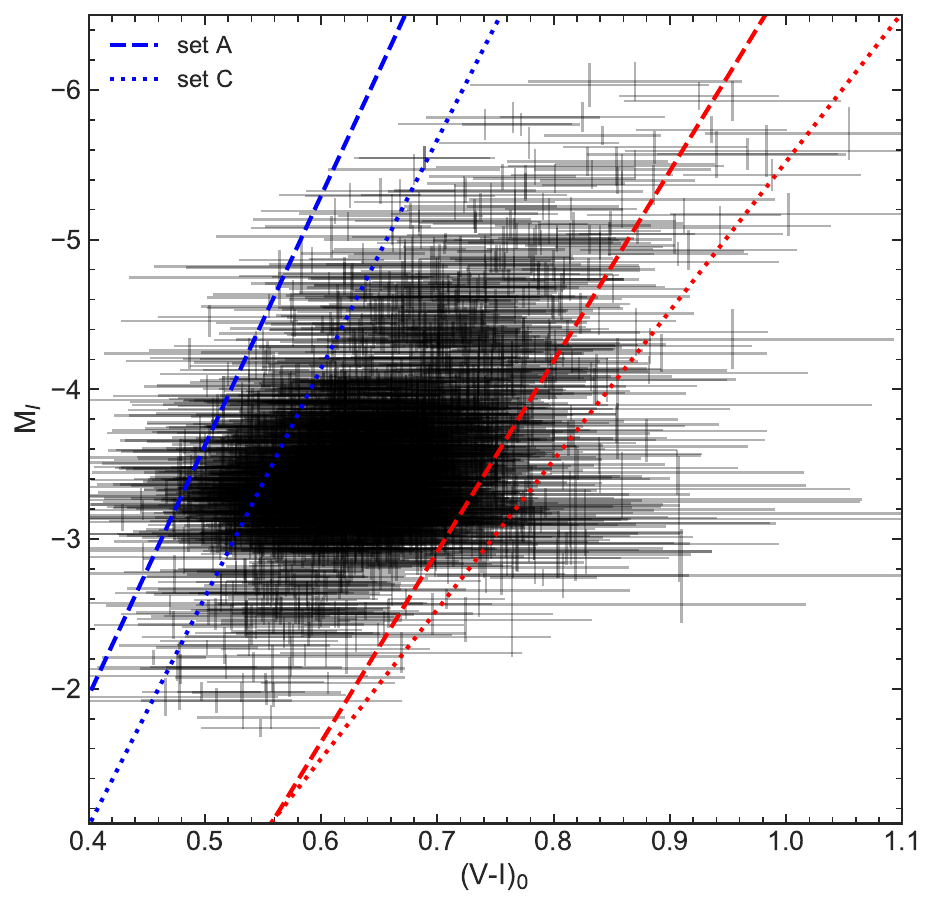}
    \caption{Color-magnitude diagram comparing the blue and red edges of the IS of the fundamental mode of pulsation for two different sets of convective parameters, namely set A (dashed lines) and set C (dotted lines). Data for fundamental-mode classical Cepheids of the LMC from the OGLE-IV catalog of variable stars are shown as dots.}
    \label{fig:fig_4}
\end{figure}

The non-adiabatic linear analysis is performed on the initial model using the linearized radial pulsation equations \citep[details in][]{Smolec2009}. These provide the eigenmodes, periods, and growth rates. The initial static model is perturbed with a linear combination of the velocity eigenmodes of the three lowest-order radial modes.

The time integration is performed for a specific number of cycles. A new cycle begins when the model passes through the maximum radius. We performed tests calculating linear and non-linear periods for $5.5$ and $7\,{\rm M}_{\odot}$ using the set C of convective parameters. In general, we found good agreement between the periods of the fundamental mode obtained in the linear analysis and after 200 cycles of time integration. The differences between linear and nonlinear periods are less than 1 per cent, which translates into differences of the same order in the PCRs. However, the effects of multi-mode pulsations are not considered in the linear periods. \citet{Paxton2019} show modeling of the Hertzsprung progression in Cepheids, produced by a resonance between the fundamental mode and a damped second overtone, using \texttt{RSP}. They found that the use of different sets of convective parameters leads to differences of, on average, 10 per cent in the nonlinear periods. Nevertheless, bump Cepheids were not included in \citet{Rodriguez2021}, and thus this source of uncertainty does not affect the comparison of our models with their data. Since integration over time for a given number of cycles for all evolutionary tracks is highly time-consuming, we consider the periods obtained in the linear non-adiabatic analysis to perform the calculation of PCRs.

\subsection{Fundamental Mode Periods}\label{subsec:Periods}

The edges of the IS for the fundamental mode were obtained with \texttt{RSP}, and are tabulated in Table~\ref{tab:table4}. These edges are sensitive to the choice of \texttt{RSP} convective parameters, as mentioned in \cite{Paxton2019}.

\begin{table}
    \centering
    \begin{tabular}{c|c|c|c|c}
    \hline\hline
        $Z$&$\alpha_{\rm blue}$&$\beta_{\rm blue}$&$\alpha_{\rm red}$&$\beta_{\rm red}$\\\hline\hline
        0.005 & -32.29&125.61&-22.27&86.24\\\hline
        0.007 & -25.86&101.09&-13.51&53.36\\\hline
        0.009 & -24.12&94.40&-13.64&53.83\\\hline\hline
    \end{tabular}
    \caption{Coefficients of the red and blue edges of the IS, assuming $\log({L/L_{\odot}})=\alpha\log{T_{\rm eff}}+\beta$, for three metallicities, namely $Z=0.005$, 0.007, and 0.009.}
    \label{tab:table4}
\end{table}

\subsubsection{Comparison of Convective Parameter Sets}\label{subsec:parameters}
In order to compare the convective parameter sets A and C, we use data of fundamental-mode classical Cepheids in the LMC from the OGLE-IV variable stars catalog \citep{Soszynski2015}, corrected for extinction using the reddening map of \cite{Skowron2020}. Data with high vertical dispersion in the reddening-free period-luminosity (PL) relation, also known as Wesenheit PL relation, were not considered. In addition, following the method described in \cite{Madore2017}, some data were discarded due to their high vertical deviation in a diagram of the magnitude residuals of the PL relation versus the corresponding residual of the Wesenheit PL relation. These deviations are possibly due to errors in the individual adopted extinctions, in addition to uncertainties due to the fact that 2D maps do not take into consideration depth-related variations in the extinction. Distances to the stars, used to calculate their absolute magnitudes, were obtained from \cite{Dobrzeniecka2016}. The edges of the IS, calculated using the convective parameter sets A and C tabulated in Table 4 of \cite{Paxton2019}, are shown in Figure~\ref{fig:fig_4}. We note that the IS of set C is bluer than that of set A, containing $87$ and $94$ per cent of the Cepheid sample, respectively. In spite of this, we adopted the convective parameters of set C since they are more physically representative. However, a combination of these sets is needed in order for the IS to contain all the classical Cepheids in the OGLE-IV catalog. This will be addressed in a future paper.

\subsubsection{Linear Periods of the Fundamental Mode}

\begin{figure*}
    \centering
    \includegraphics[scale=0.48]{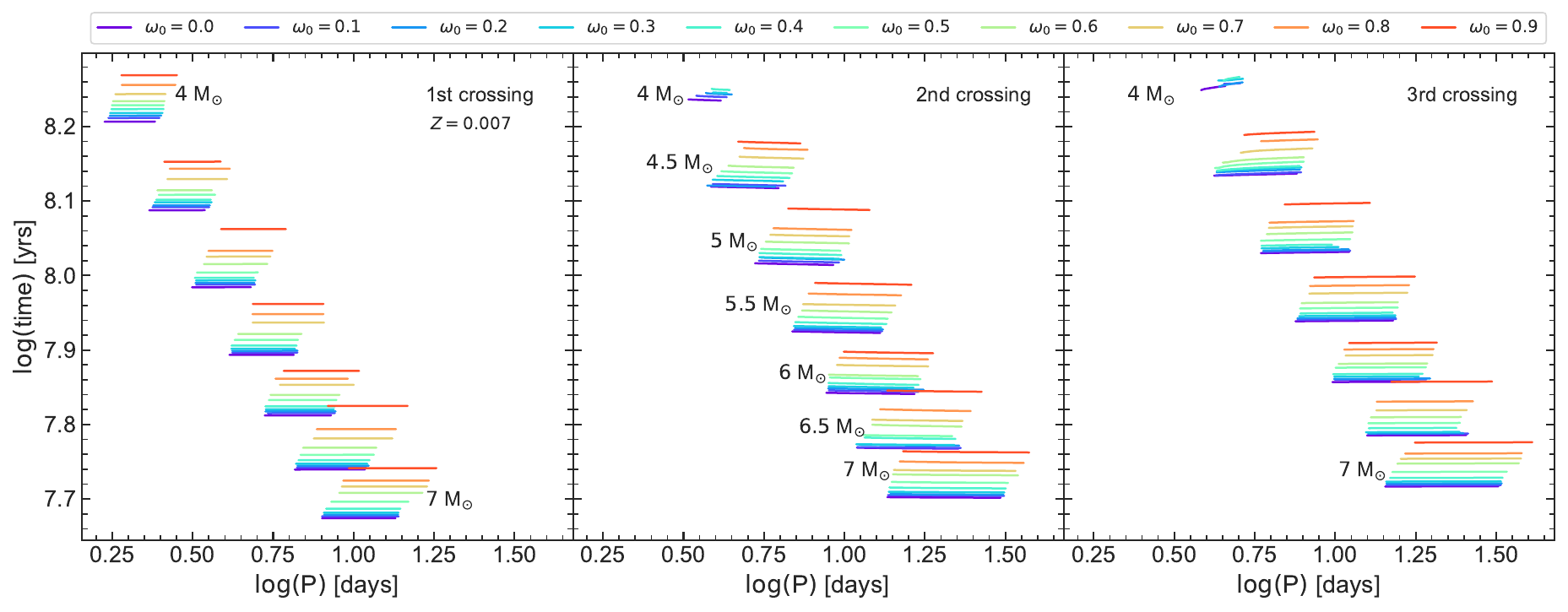}
    \caption{Time as a function of the linear periods obtained with RSP. Periods are shown for the first crossing (left panel), second crossing (middle panel), and third crossing (right panel). Periods for masses, from top to bottom, from $4$ to $7 \, {\rm M}_{\odot}$ in steps of $0.5 \, {\rm M}_{\odot}$ are displayed in each panel. Lines of different colors correspond to different $\omega_0$, following the color scheme given at the top of this figure.}
    \label{fig:fig4_1}
\end{figure*}

Linear periods for the fundamental mode  were calculated for each evolutionary track crossing the IS. These are shown in Figure~\ref{fig:fig4_1}. The time scales are of the order of $\sim0.02$, 0.4, and 0.7~Myr for the first, second, and third crossing, respectively. As mentioned in previous sections, during the first and third crossings the period increases, while during the second crossing the period decreases. For $\omega_0>0.5$, the blue loops of tracks with $4\,{\rm M}_{\odot}$ are outside the IS, as shown in Figure~\ref{fig:fig3_2}, hence fundamental-mode pulsations are not excited in these models. It can also be noted that, as the rotation rate increases, a Cepheid of a given mass tends to reach a slightly longer period. This can be understood in terms of the PL relation (Leavitt's law): since rotating tracks are more luminous than non-rotating tracks, the latter can be expected to display shorter periods. Note that the period ranges that we obtain are in agreement with those shown in Figure 5 of \cite{Anderson2016}.

\subsubsection{Period-Age and Period-Age-Temperature Relations}

Cepheids obey period-age (PA) and period-age-color (PAC) relations \citep[see, e.g.,][]{Kippenhahn1969,Bono2005}. Considering the mass-luminosity relation \citep{Eddington1924}, stellar ages from evolutionary models, and PL relations, Cepheids with a high initial mass, a high luminosity, and a long period are younger than low-mass Cepheids with lower luminosity and shorter periods. In addition, at a given luminosity, the period increases toward the red edge of the IS. 

\begin{figure*}
    \centering
    \includegraphics[scale=0.48]{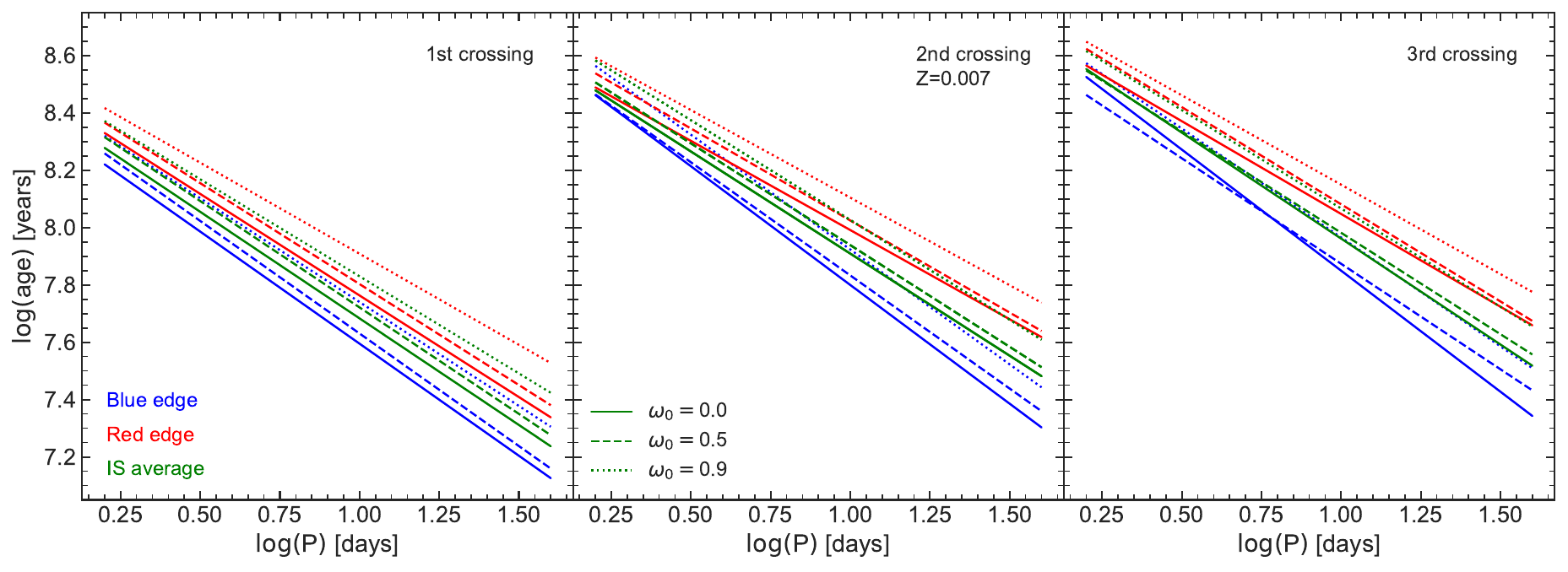}
    \includegraphics[scale=0.48]{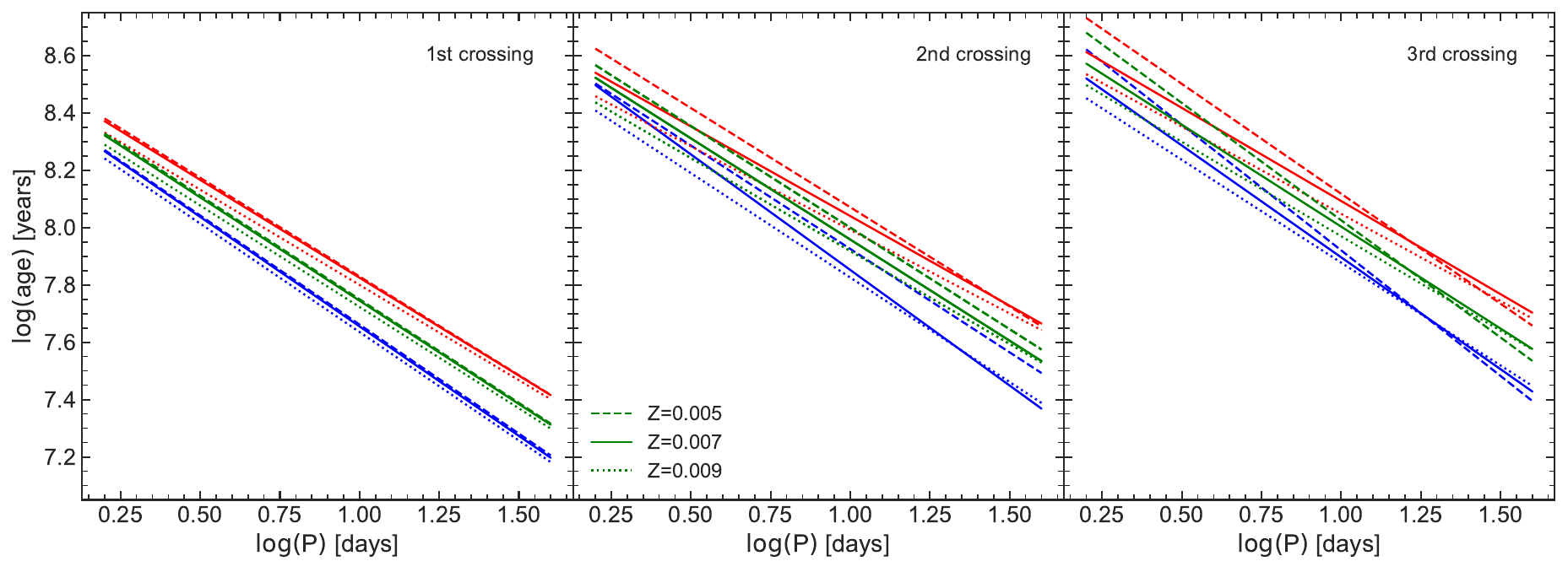}
    \caption{PA relationships obtained from the linear periods calculated with RSP, for the first crossing (left panels), second crossing (middle panels), and third crossing (right panels). The upper row shows the relationships for three initial rotation rates, namely $\omega_0 = 0.0$ (solid line), $0.5$ (dashed line), and $0.9$ (dotted line), considering $Z=0.007$. The lower row shows the relationships for three different metallicities, namely $Z=0.005$ (dashed line), $0.007$ (solid line), and $0.009$ (dotted line). In each panel, the PA relationship for the blue (blue lines) and red (red lines) edges of the IS, as well as for an average over the IS (green lines), is shown.}
    \label{fig:fig4_3}
\end{figure*}

We derive PA relations by performing ordinary least-squares regression with the obtained periods and evolutionary ages;
these are shown in Figure~\ref{fig:fig4_3}. We show relations for $Z=0.007$ with $\omega_0=0.0$, $0.5$, $0.9$, and relations averaged over the different assumed initial rotation rates for metallicities $Z=0.005$, $0.007$, and $0.009$, respectively. We note dependencies on four important parameters: crossing number, since the three crossings occur in sequence and the time scales of the first crossing are much shorter than those of the second and third crossings, resulting in slightly steeper relations; position in the IS, due to the change of the pulsation period produced by the expansion or contraction of the star, producing appreciable changes in slope and zero point; rotation, since these models have a longer MS due to rotational mixing, causing a significant change in age in fast-rotating stars; and metallicity, which produces appreciable changes in the age of the star for short pulsation periods, similar to the effect produced by rotation in stars with longer periods. The PA relations obtained for the red and blue fundamental edges, in addition to an average relation, are tabulated in Table~\ref{tab:table4_1}. We also include period-age-temperature (PAT) relations, where the temperature is a proxy for the color of the star. The latter relations take into account the temperature range of the IS, resulting in a better representation of the age of the star. In general, the residuals of these relationships are small, less than $0.04$~dex in log-age.

\begin{figure*}
    \centering
    \includegraphics[scale=0.65]{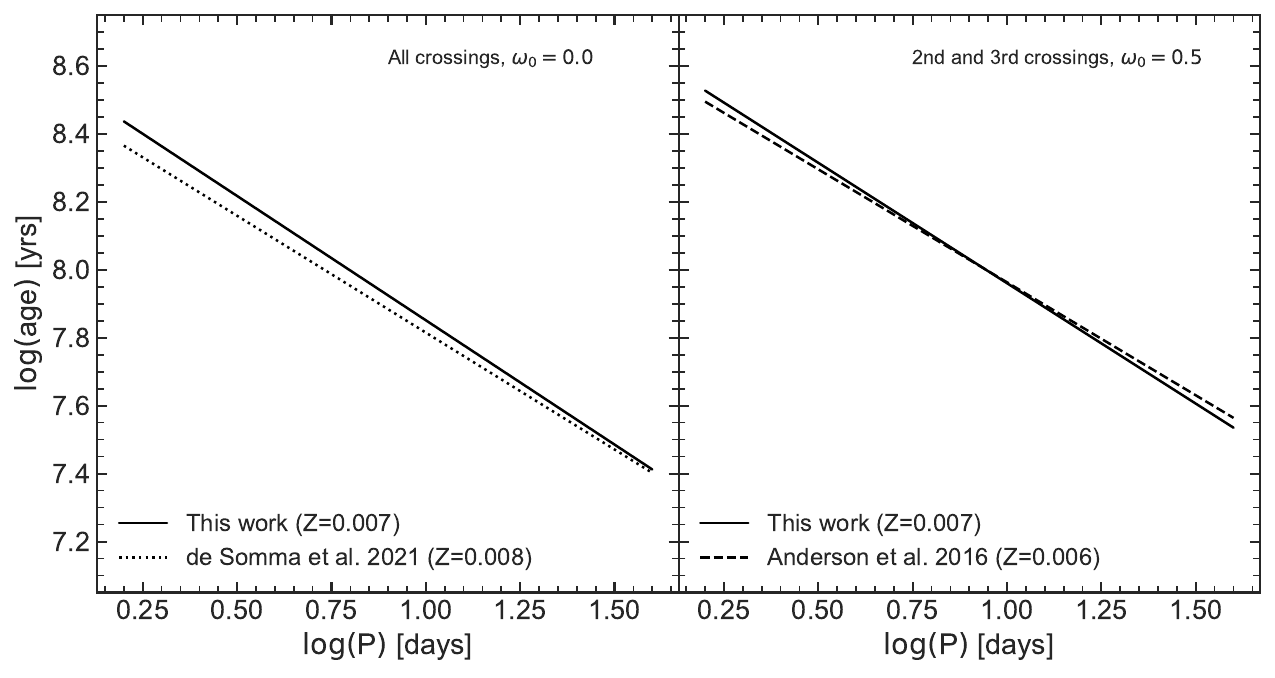}
    \caption{Comparison between the PA relations from this work (solid line) and those presented by \citet[][dotted line, left panel]{DeSomma2021} and \citet[][dashed line, right panel]{Anderson2016}. The left panel shows the PA relationships averaged over all IS crossings, and without rotation. The right panel shows the PA relations averaged over the second and third IS crossings, and an initial rotation rate of $\omega_{0}=0.5$. The coefficients of our PA relations shown in this figure can be found in Table~\ref{tab:table4_1}. See text for further details.}
    \label{fig:compasiron_PA}
\end{figure*}

In Figure~\ref{fig:compasiron_PA}, we compare our PA relation with those computed by \citet{Anderson2016} and \citet[][their ``case B'']{DeSomma2021}. In order to do this, we adopt similar assumptions as in those studies, choosing $\omega_{0}=0.0$ and averaging over all three IS crossings in the case of \citet{DeSomma2021}, and adopting $\omega_{0}=0.5$ and averaging over the second and third crossings only in the case of \citet{Anderson2016}. Under these assumptions, we find reasonable agreement with the PA relations of both these authors. Note, in this sense, that the \citet{Anderson2016} relation was calculated assuming instantaneous overshooting with an overshooting parameter of $f_{\rm ov}=0.1$, as discussed in Section~\ref{subsec:anderson}, and a metallicity $Z=0.006$. ``Case B'' models from \citet{DeSomma2021}, in turn, are based on Bag of Stellar Tracks and Isochrones \citep[BaSTI;][]{Hidalgo2018} evolutionary tracks that also take into account core overshooting and mass loss, but not rotation, and were computed for a metallicity $Z = 0.008$. As far as their pulsation calculations go, \citet{DeSomma2021} approximate the effects of overshooting by considering an increase in the luminosity, over their canonical models, of $\Delta(\log L/L_{\sun}) = 0.2$~dex. As can be seen in Figure~\ref{fig:compasiron_PA}, this leads to a PA relation that is only slightly offset from our corresponding one, in the sense that the \citet{DeSomma2021} relation predicts lower ages for a given period, the relative difference increasing towards shorter periods/younger ages.

\begin{table}
    \centering
    \begin{tabular}{c|c|c|c|c|c|c|c}
        \hline
        $\omega_0$&Crossing&$\alpha_{\rm blue}$&$\beta_{\rm blue}$&$\alpha_{\rm red }$&$\beta_{\rm red}$&$\alpha$&$\beta$\\\hline
        0.0 &1st& -0.78&8.38&-0.71&8.47&-0.74&8.43\\
            &2nd& -0.83&8.63&-0.62&8.61&-0.71&8.62\\
            &3rd& -0.85&8.70&-0.65&8.70&-0.74&8.70\\
            &all& & & & & -0.73 & 8.58 \\\hline
        0.5 &1st& -0.79&8.42&-0.71&8.51&-0.74&8.46\\
            &2nd& -0.80&8.62&-0.64&8.67&-0.71&8.65\\
            &3rd& -0.74&8.61&-0.68&8.76&-0.71&8.69\\
            &2nd and 3rd& & & & & -0.71 & 8.67 \\\hline
        0.9 &1st& -0.73&8.47&-0.64&8.54&-0.68&8.51\\
            &2nd& -0.80&8.73&-0.61&8.72&-0.70&8.72\\
            &3rd& -0.76&8.73&-0.62&8.77&-0.69&8.75\\\hline
        $Z$&Crossing&$\alpha_{\rm blue}$&$\beta_{\rm blue}$&$\alpha_{\rm red}$&$\beta_{\rm red}$&$\alpha$&$\beta$\\\hline
        0.005 &1st& -0.82&8.45&-0.74&8.56&-0.78&8.51\\
            &2nd&   -0.72&8.65&-0.69&8.76&-0.71&8.71\\
            &3rd&   -0.88&8.80&-0.77&8.89&-0.82&8.84\\\hline
        0.007 &1st& -0.76&8.42&-0.68&8.51&-0.72&8.47\\
            &2nd&   -0.81&8.66&-0.63&8.67&-0.71&8.66\\
            &3rd&   -0.78&8.68&-0.65&8.74&-0.71&8.72\\\hline
        0.009 &1st& -0.72&8.37&-0.64&8.44&-0.68&8.40\\
            &2nd&   -0.73&8.55&-0.58&8.58&-0.65&8.57\\
            &3rd&   -0.72&8.60&-0.61&8.66&-0.66&8.63\\\hline
    \end{tabular}
    \caption{PA relation coefficients, assuming $\log({\rm age}
    /{\rm year})=\alpha\log(P/{\rm days}) + \beta$. The top four rows correspond to the blue edge ($\alpha_{\rm blue}$ and $\beta_{\rm blue}$), red edge ($\alpha_{\rm red}$ and $\beta_{\rm red}$), and an average over the IS ($\alpha$ and $\beta$), calculated for the three initial rotation rates that are shown in column~1, for three crossings of the IS as shown in column~2, and $Z=0.007$. The lower four rows correspond to the same coefficients as calculated for the three metallicities shown in column~1, averaged over rotation. In addition, in order to facilitate comparison with previous work by \citet{Anderson2016} and \citet{DeSomma2021}, for $\omega_0=0.0$ and $0.5$ we also include PA relation coefficients averaged over all crossings and averaged over only the second and third crossings of the IS, respectively.}
    \label{tab:table4_1}
\end{table}

\begin{table}
    \centering
    \begin{tabular}{c|c|c|c|c}
            \hline
        $Z$&Crossing&$\alpha$&$\beta$&$\gamma$\\\hline
        0.005   &1st&-0.87&-3.0&19.84\\
                &2nd&-0.92&-3.39&21.57\\
                &3rd&-0.98&-3.87&23.45\\\hline
        $Z$&Crossing&$\alpha$&$\beta$&$\gamma$\\\hline
        0.007   &1st&-0.90&-3.03&19.96\\
                &2nd&-0.91&-3.51&21.95\\
                &3rd&-0.92&-3.58&22.27\\\hline
        $Z$&Crossing&$\alpha$&$\beta$&$\gamma$\\\hline
        0.009   &1st&-0.88&-2.88&19.35\\
                &2nd&-0.85&-3.16&20.57\\
                &3rd&-0.87&-3.29&21.1\\\hline
    \end{tabular}
    \caption{Coefficients of the 
    PAT relation for three different metallicities and for the three noted crossings of the IS,  averaged over rotation and assuming $\log({\rm age}/{\rm year})=\alpha\log{(P/{\rm days})} + \beta\log{(T_{{\rm eff}}/{\rm K})}+\gamma$. }
    \label{tab:table4_2}
\end{table}
\subsubsection{Period-Luminosity and Period-Luminosity-Temperature Relations}

\begin{figure*}
    \centering
    \includegraphics[scale=0.48]{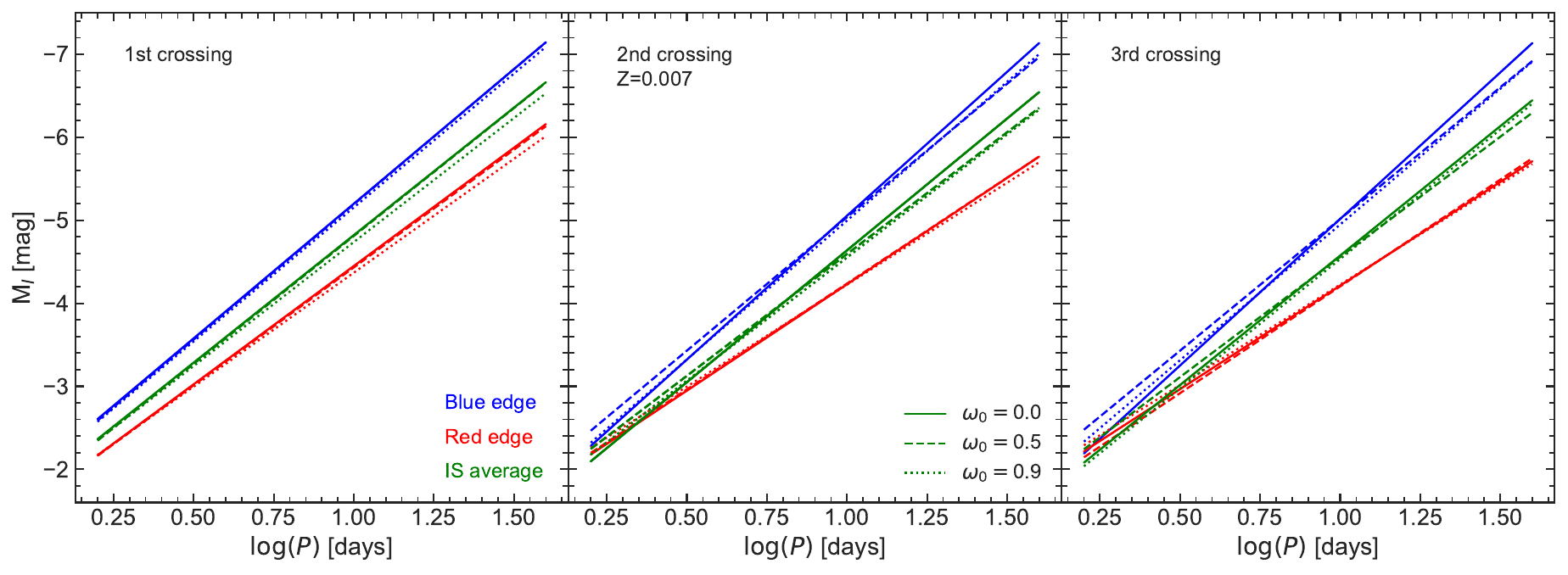}
    \includegraphics[scale=0.48]{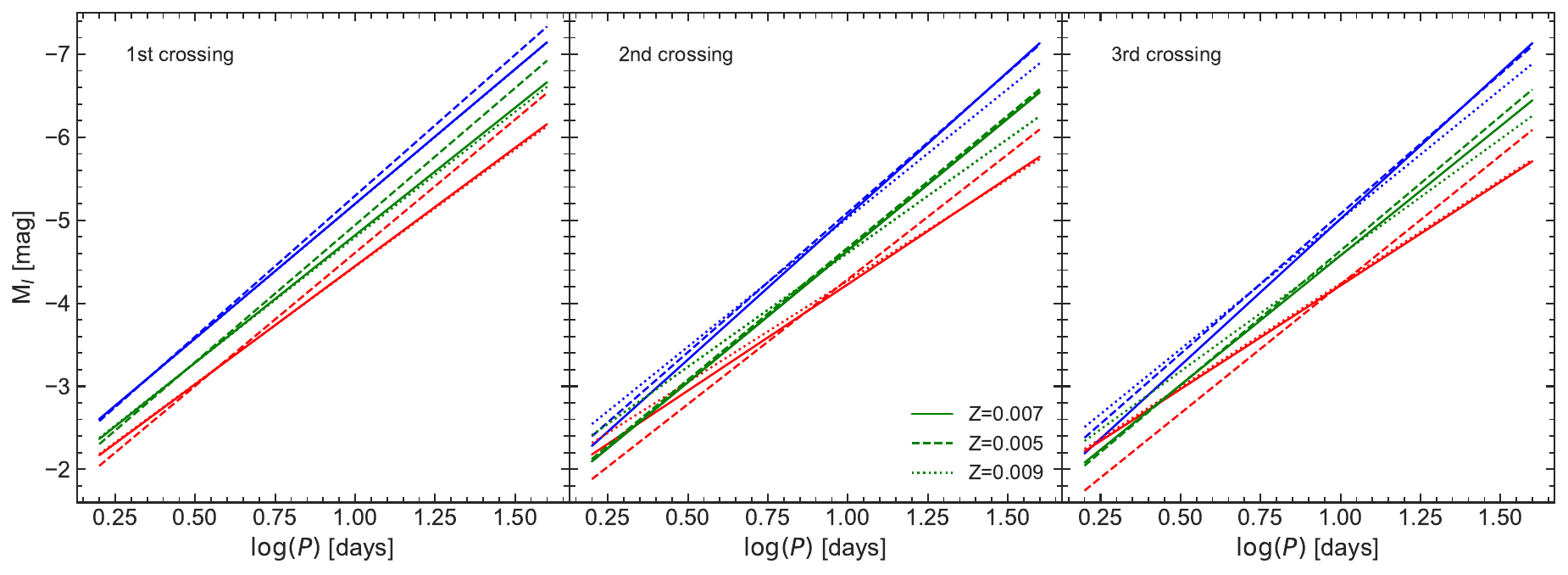}
    \caption{PL relationships obtained from the linear periods calculated with RSP. This relationship was calculated for the first crossing (left panels), second crossing (middle panels) and third crossing (right panels). The upper panels show the relationships for three initial rotation rates $\omega_0 = 0.0$ (solid line), $0.5$ (dashed line) and $0.9$ (dotted line), considering Z=0.007. The lower panels show the relationships for three different metallicities, namely $Z=0.005$, $0.007$ and $0.009$. In each panel, PL relationship for the blue (blue lines), red (red lines) edge of the IS, and an average (green lines) in the IS are shown.}
    \label{fig:fig4_5}
\end{figure*}

In Figure~\ref{fig:fig4_5}, PL relations for the \textit{I} band are shown. In the upper panels, we compare the effects of rotation, while in the lower panels we compare the effects of metallicity on non-rotating models. These relationships have slightly larger residuals than the PA and PAT relationships, of the order of $0.1$~dex in magnitude. We note that rotation tends to slightly broaden the PL relation, both for short and long periods, but the broadening is minimal for periods close to 10 days. This complex behavior is due to the non-monotonic relationship between rotation and luminosity, as discussed in Sect.~\ref{sec:HDR}. 
 
On the other hand, considering a wider range in metallicity in the fit broadens it further, although it can be noted that this broadening is minimal during the first crossing of the IS.

\begin{figure*}
    \centering
    \includegraphics[scale=0.48]{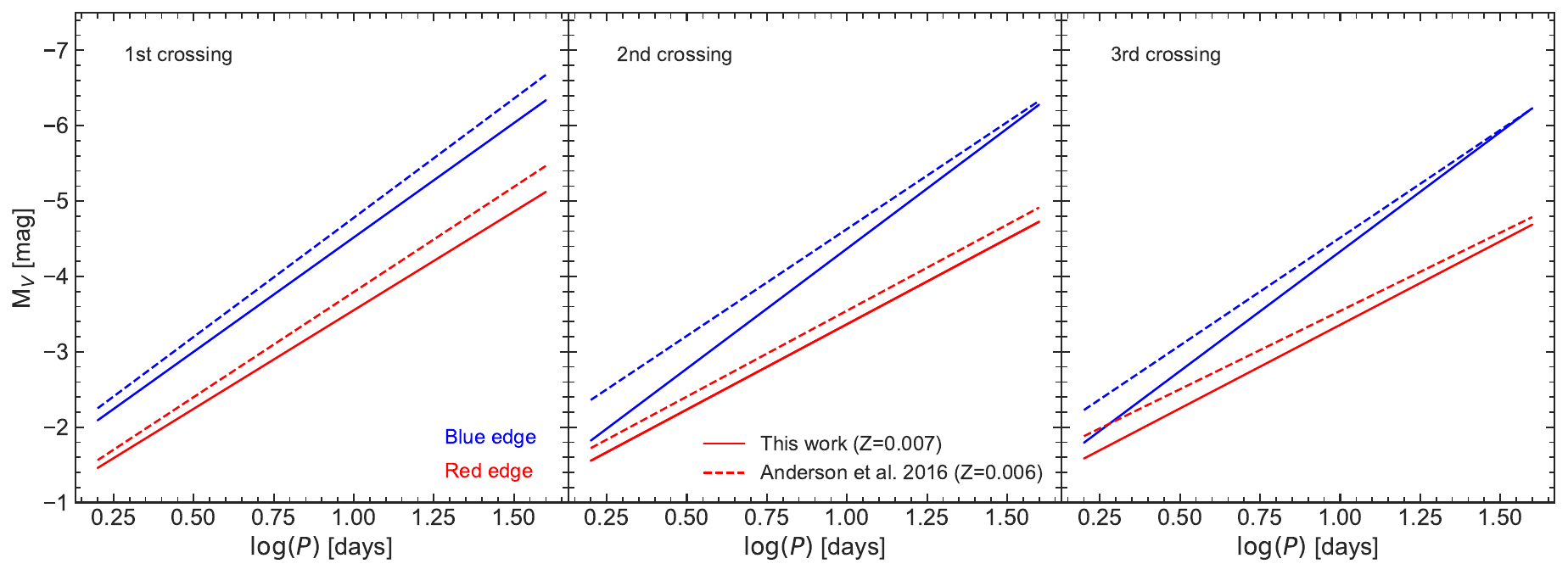}
    \caption{Comparison of PL relations of this work (solid lines) with those presented by \citet[][dashed lines]{Anderson2016}, for the first (left panel), second (central panel), and third (right panel) IS crossing. In each panel, the PL relation for the blue (blue lines) and red (red lines) IS edges are shown.}
    \label{fig:comparison_PL}
\end{figure*}

A comparison of our PL relation with those of \citet{Anderson2016} in the \textit{V} band is shown in Figure~\ref{fig:comparison_PL}, for the three IS crossings and averaged over three initial rotation rates, namely $\omega_0 = 0.0, 0.5$ and $0.9$. We note that the PL relations for the blue and red edges of Anderson's models are systematically brighter than those obtained in this work.

In the same way as for the PAT relations, we tabulate the coefficients of the period-luminosity-temperature (PLT) relations in Table~\ref{tab:table4_4}.

\begin{table}
    \centering
    \begin{tabular}{c|c|c|c|c|c|c|c}
        \hline
        $\omega$&Crossing&$\alpha_{\rm blue}$&$\beta_{\rm blue}$&$\alpha_{\rm red}$&$\beta_{\rm red}$&$\alpha$&$\beta$\\\hline
        0.0 &1st& -3.24&-1.96&-2.85&-1.60&-3.07&-1.75\\
            &2nd& -3.47&-1.59&-2.56&-1.67&-3.18&-1.46\\
            &3rd& -3.53&-1.49&-2.50&-1.72&-3.12&-1.46\\\hline
        0.5 &1st& -3.26&-1.94&-2.84&-1.60&-3.08&-1.73\\
            &2nd& -3.21&-1.82&-2.55&-1.69&-2.93&-1.66\\
            &3rd& -3.17&-1.84&-2.57&-1.63&-2.90&-1.66\\\hline
        0.9 &1st& -3.23&-1.93&-2.74&-1.62&-2.99&-1.75\\
            &2nd& -3.34&-1.65&-2.46&-1.77&-2.97&-1.58\\
            &3rd& -3.27&-1.68&-2.42&-1.81&-3.12&-1.41\\\hline
        $Z$&Crossing&$\alpha_{\rm blue}$&$\beta_{\rm blue}$&$\alpha_{\rm red}$&$\beta_{\rm red}$&$\alpha$&$\beta$\\\hline
        0.005 &1st& -3.32&-1.91&-3.00&-1.49&-3.15&-1.69\\
            &2nd&   -2.97&-1.88&-2.79&-1.43&-2.88&-1.64\\
            &3rd&   -3.41&-1.52&-2.91&-1.28&-3.15&-1.39\\\hline
        0.007 &1st& -3.24&-1.94&-2.81&-1.61&-3.05&-1.74\\
            &2nd&   -3.34&-1.69&-2.52&-1.71&-3.03&-1.57\\
            &3rd&   -3.32&-1.67&-2.50&-1.72&-3.04&-1.51\\\hline
        0.009 &1st& -3.20&-1.97&-2.66&-1.75&-2.91&-1.85\\
            &2nd&   -3.15&-1.86&-2.43&-1.83&-2.75&-1.84\\
            &3rd&   -3.14&-1.84&-2.39&-1.89&-2.74&-1.85\\\hline
    \end{tabular}
    \caption{\textit{I}-band PL relation (in the form $M_{I}=\alpha\log(P/{\rm days}) + \beta$) coefficients. The top four rows correspond to coefficients for the blue edges ($\alpha_{\rm blue}$ and $\beta_{\rm blue}$), red edges ($\alpha_{\rm red}$ and $\beta_{\rm red}$), and an average over the IS ($\alpha$ and $\beta$), calculated for three initial rotation rates ($\omega_0 = 0.0$, $0.5$, and $0.9$), for three crossings of the IS, and $Z=0.007$. The bottom four rows correspond to these same coefficients, but calculated for three metallicities, namely $Z=0.005$, $0.007$, and $0.009$, averaged over rotation.}
    \label{tab:table4_3}
\end{table}

\begin{table}
    \centering
    \begin{tabular}{c|c|c|c|c}
            \hline
        $Z$&Crossing&$\alpha$&$\beta$&$\gamma$\\\hline
        0.005   &1st&-3.71&-12.63&46.08\\
                &2nd&-3.78&-14.50&53.31\\
                &3rd&-3.79&-14.60&53.72\\ \hline
        $Z$&Crossing&$\alpha$&$\beta$&$\gamma$\\\hline
        0.007   &1st&-3.80&-12.85&46.95\\
                &2nd&-3.73&-14.71&54.02\\
                &3rd&-3.73&-14.85&54.56\\ \hline
        $Z$&Crossing&$\alpha$&$\beta$&$\gamma$\\\hline
        0.009   &1st&-3.78&-12.74&46.50\\
                &2nd&-3.68&-14.31&52.45\\
                &3rd&-3.67&-14.48&53.10\\\hline
    \end{tabular}
    \caption{PLT relation (in the form $M_{I}=\alpha\log(P/{\rm days}) + \beta\log{(T_{\rm eff}/{\rm K})}+\gamma$) coefficients, for $Z=0.007$, averaged over rotation and for each crossing of the IS.}
    \label{tab:table4_4}
\end{table}

\subsection{Period Change Rates}\label{sec:pdot}

\begin{figure*}
    \centering
    \includegraphics[scale=0.47]{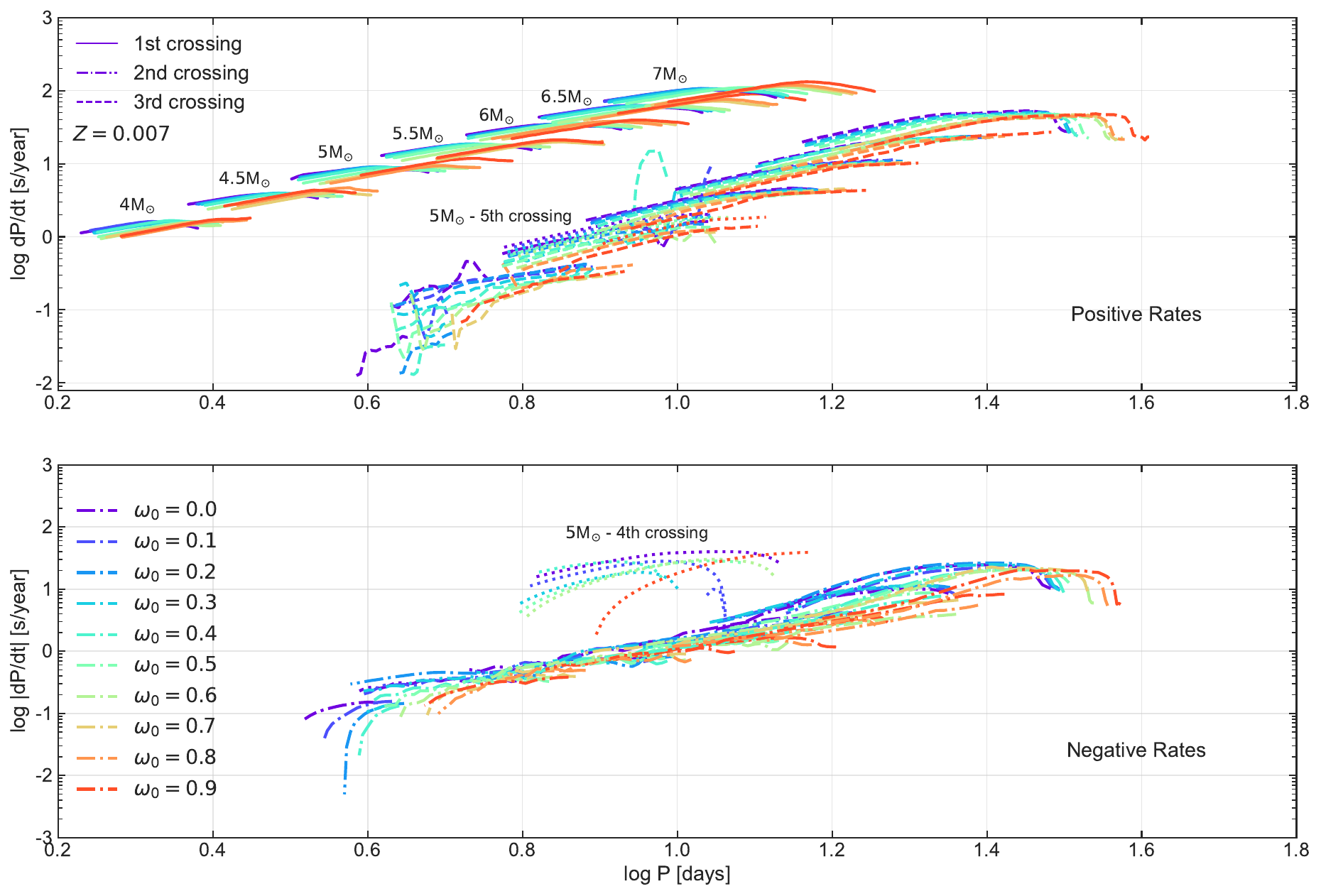}
        \caption{PCRs as a function of the period for evolutionary tracks with $Z=0.007$. The upper panel shows positive rates corresponding to the first (solid lines), third (dashed lines) and fifth (dotted lines) crossings of the IS. The lower panel shows negative rates corresponding to the second (dashed-dotted lines) and fourth (dotted lines) crossing of the IS. For each panel, PCRs are shown for masses from left to right, from $4$ to $7 \, {\rm M}_{\odot}$ in steps of $0.5 \, {\rm M}_{\odot}$. Lines of different colors correspond to different $\omega_0$, following the color scheme given at the left of the lower panel.}
    \label{fig:fig_dpdt_4to5to}
\end{figure*}

Evolutionary PCR values were calculated directly from the linear periods obtained with RSP, and are shown in Figure~\ref{fig:fig_dpdt_4to5to}. The behavior of the blue loops in the HRD defines the shapes of the PCR vs. period curves; in the case of stars of $4\, {\rm M}_{\odot}$, the blue loops showed a complex behavior due to a large number of He-spikes, which is reflected in the curves of the second and third crossings, where the PCRs show a behavior with a similar degree of complexity. We note a dependence of the PCR on rotation. First-crossing non-rotating models tend to have a higher PCR at the beginning of the IS than rotating models, but the models with the highest initial rotation rates are those with the highest PCRs at the red end of the IS. For the first and third crossings, in most cases, non-rotating tracks have a higher PCR during the entire crossing of the IS, whereas the dependence on rotation is not monotonic in the case of the second crossing.

For $5 \, {\rm M}_{\odot}$ stars with $\omega_0=0.4$ at the third crossing, a small bump in luminosity takes place that leads to a large increase in PCR values, which reach the area of the first crossing. In addition, in Figure~\ref{fig:fig_dpdt_4to5to} we add dotted lines corresponding to the PCRs of the fourth and fifth crossings of $5 \, {\rm M}_{\odot}$ stars. The PCR of the fourth crossing is negative, but it is larger by more than an order of magnitude, in absolute value, than the rate of the second crossing. In contrast, the rate of the fifth crossing is positive and is blended in the figure with the rate of the third crossing. The typical duration of the fourth crossing is roughly twice the timescale of the first crossing of the IS, while the duration of the fifth crossing is comparable to that of the third crossing of the IS. If irregularities in the evolutionary paths of Cepheids across the HRD, such as those produced by a fourth or fifth crossing of the IS, are real, perhaps attributed to mixing episodes that may inject a substantial amount of helium into the core, they may produce potentially observable effects in PCRs, in spite of the very short duration of these events.

\subsection{Comparison with Empirical PCRs}\label{sec:comparison}

Cepheids are dubbed ``magnifying glasses of stellar evolution'' \citep{Kippenhahn1994} because they provide a highly sensitive test of it. Therefore, it is essential to compare theoretical results with empirical data for actual Cepheids, in order to gain insight into the adequacy of the input physics that is used to build these models. Of particular interest is comparing the results obtained by us and previous authors for classical Cepheids in the LMC, given the large amount of data currently available. In our previous paper \citep{Rodriguez2021}, we have computed PCRs for an unprecedented number of LMC Cepheids, making it a key work for comparison with model predictions.

\begin{figure*}
    \centering
    \includegraphics[scale=0.47]{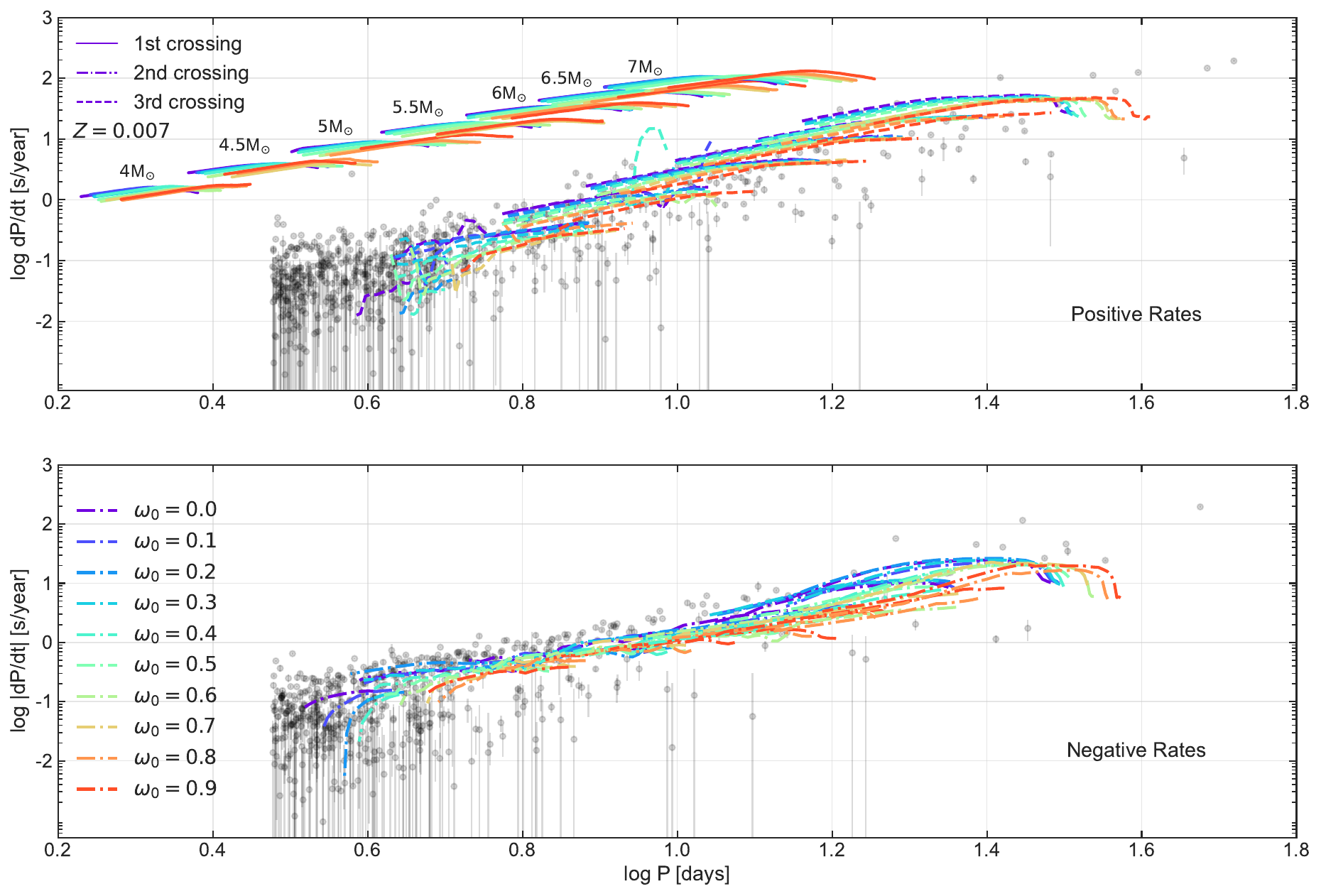}
    \caption{Comparison between the PCRs for models with $Z=0.007$ obtained in this work and the empirical rates calculated by \protect\cite{Rodriguez2021} for LMC Cepheids. The upper panel shows positive rates corresponding to the first (solid lines) and third (dashed lines) crossings of the IS. The lower panel shows negative rates corresponding to the second crossing of the IS (dashed-dotted lines). In each panel, PCRs are shown for masses, from left to right, from $4$ to $7 \, {\rm M}_{\odot}$, in steps of $0.5 \, {\rm M}_{\odot}$. Lines of different colors correspond to different $\omega_0$ values, following the color scheme given at the left of the lower panel.}
    \label{fig:fig_dpdt_data}
\end{figure*}

\begin{figure*}
    \centering
    \includegraphics[scale=0.47]{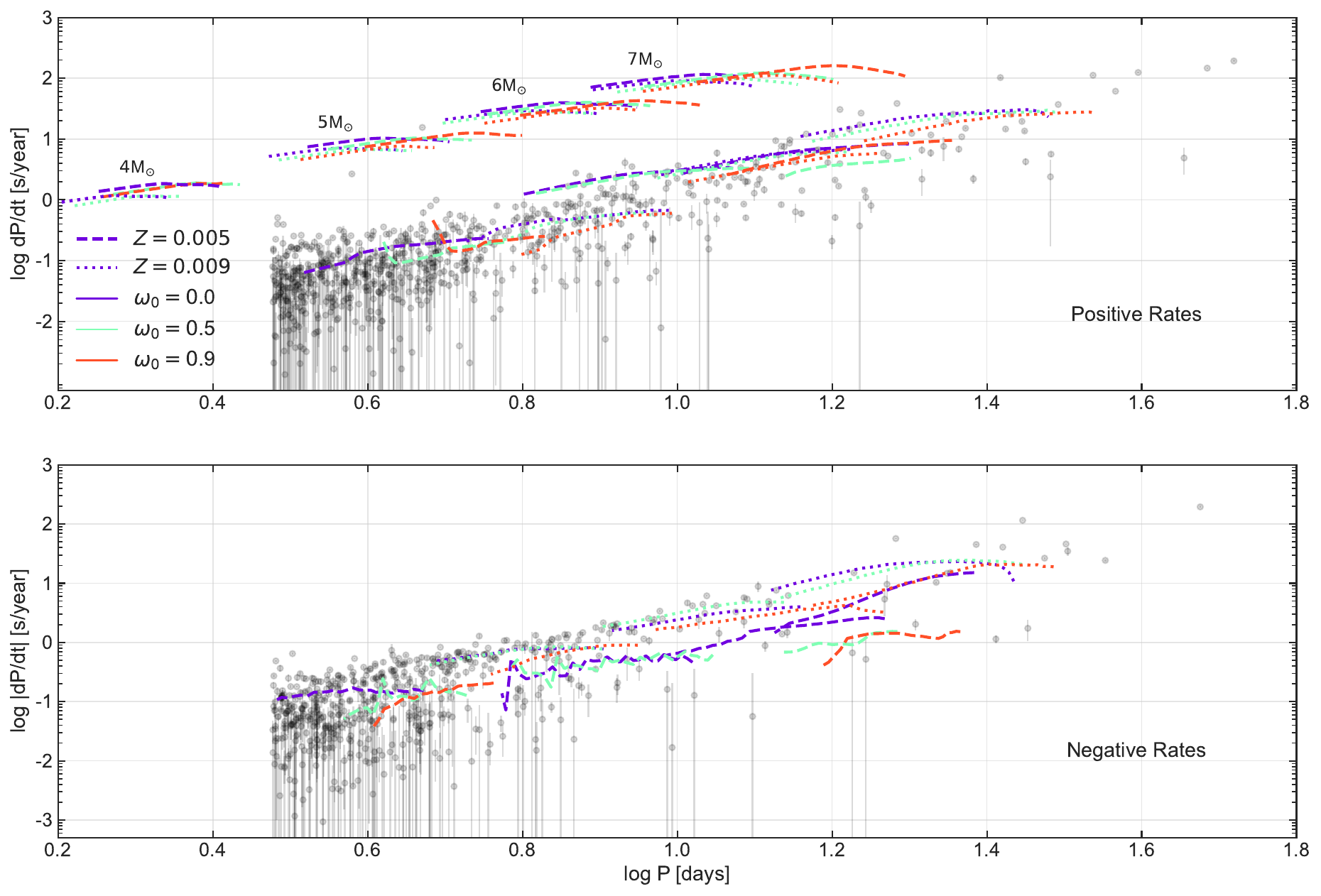}
    \caption{
   Comparison between the PCRs for models with $Z=0.005$ (dashed lines) and $Z=0.009$ (dotted lines) obtained in this work and the empirical rates calculated by \protect\cite{Rodriguez2021} for LMC Cepheids. The upper panel shows positive rates corresponding to the first (upper left part) and third (lower right part) crossings of the IS. The lower panel shows negative rates corresponding to the second crossing of the IS. For each panel, PCRs are shown for masses, from left to right, from $4$ to $7 \, {\rm M}_{\odot}$, in steps of $1 \, {\rm M}_{\odot}$. Lines of different colors correspond to different $\omega_0$ values, following the color scheme given at the  bottom left of the upper panel.}
    \label{fig:fig_dpdt_data_Zs}
\end{figure*}

Figure~\ref{fig:fig_dpdt_data} compares our PCRs for $Z=0.007$ and the empirical values obtained in \cite{Rodriguez2021}. A similar comparison is made in Figure~\ref{fig:fig_dpdt_data_Zs} for metallicities $Z=0.005$ and $Z=0.009$. There is good general agreement between the models and the data, but there are several points to note. Most of the data sample are classical Cepheids with implied masses near or below $4\,{\rm M}_{\odot}$, with a wide range of possible initial rotation rates. On the other hand, in the long-period region, a couple of Cepheids with implied masses greater than $7\,{\rm M}_{\odot}$ are observed, which makes them interesting sources since the vast majority of Cepheid masses measured to date lie between $3.6$ and $5\,{\rm M}_{\odot}$ \citep{Pilecki2018,Evans2018,Gallenne2019}. In the area of positive rates and short periods of Figure~\ref{fig:fig_dpdt_data}, models with $Z=0.007$ do not cover the data for the shortest-period Cepheids. Models with the same metallicity but masses lower than $4\,{\rm M}_{\odot}$ could in principle cover this low-period regime. However, with the adopted input physics, we were unable to produce models with $< 4\,{\rm M}_{\odot}$ that crossed the IS during core He burning, as these produced blue loops that were too short, similarly to what was previously found by \citet{Anderson2016}. Models with different metallicity cover a wider range of periods. In the negative rate panel of Figures \ref{fig:fig_dpdt_data} and \ref{fig:fig_dpdt_data_Zs}, the PCRs of the $4\,{\rm M}_{\odot}$ model do cover the area of the shortest periods. In addition, several Cepheids have smaller or larger PCRs than the main loci predicted by our models. A complication that follows from Figures~\ref{fig:fig_dpdt_data} and \ref{fig:fig_dpdt_data_Zs} is that there is a degeneracy in these models, since a curve with a certain mass, metallicity, and initial rotation rate can overlap a curve with a different combination of these parameters. A wider grid in metallicity and masses is needed to cover the full range of the \cite{Rodriguez2021} data. 

\section{Summary and Conclusions}\label{sec:conclusions}

In this work, we use the MESA stellar evolution code to calculate PCRs for a set of evolutionary models of intermediate-mass stars. These models span a mass range between $4$ and $7\, {\rm M}_{\odot}$; metallicities of $Z=0.005$, $0.007$, and $0.009$; and an initial rotation rate between $\omega_0=0.0 - 0.9$, in steps of 0.1.

Rotation has a major impact on the evolution of a star. Due to the competition between centrifugal force and rotational mixing, the duration of the MS is extended by $15.4$ per cent, on average. During the helium-burning phase, in the blue loop, we notice an increase in luminosity as rotation increases. Moreover, for stars with $M\leq5.5 \, {\rm M}_{\odot}$, the extent of the blue loop decreases as the rotation rate increases. On the other hand, for stars with $M> 5.5 \, {\rm M}_{\odot}$, the behavior of the blue loop extension is non-monotonic. In addition, multiple ``He-spikes'' are observed, especially in the tracks between $4$ and $5\,{\rm M}_{\odot}$.

We use RSP to obtain linear periods at each IS crossing for each stellar track. In addition, the IS edges were also computed using RSP. We observe dependencies with RSP convective parameters. Linear periods for the fundamental mode were also obtained along each evolutionary track using RSP. PA, PAT, PL, and PLT relationships were computed for three rotation rates and metallicities. The PA relationships showed dependencies on crossing number, position in the IS, rotation, and metallicity. The PL relations become slightly broader with increasing rotation at the second and third crossings. Our models also reveal that additional broadening is brought about by metallicity variations.

PCRs were calculated directly from the linear periods. The behavior of the blue loops in the HRD defines the shapes of the PCR vs. period diagram. First crossing non-rotating models show a higher PCR at the blue end of the IS, while models with a high rotation rate show a higher PCR at the red end of the IS. Models of the second crossing show a non-monotonic dependence on rotation. On the other hand, non-rotating tracks show a higher PCR during the third crossing. 

We compared our models with those presented in \cite{Anderson2016}. For both $5 \,{\rm M}_{\odot}$ and $7 \,{\rm M}_{\odot}$ models, major differences are found between the properties of our tracks and the corresponding ones presented by \cite{Anderson2016}, including ZAMS position, MS extent, RGB temperature, and blue loop behavior. Some of these differences are attributed to differences in the adopted physical parameters, such as mixing length and overshooting. However, the differences in the rotation implementation are the most important when comparing the evolutionary tracks. \cite{Anderson2016} presented PCRs for models with $Z=0.006$. When compared to our values, the PCRs are of the same order of magnitude as the values obtained in this work, for models with $Z=0.007$ and $Z=0.005$.

\cite{Rodriguez2021} recently measured PCRs for a sample of 1303 LMC classical Cepheids. Comparing their results with our models in the PCR vs. period diagram, good general agreement is observed. However, our models do not cover the short-period regime, and present a lower PCR values than those observed in the data. The inability of our models to cover the short-period region has two main reasons. First, the blue edge of the IS, calculated with RSP, is too cold in comparison with the blue edge that is implied by the observed Cepheid colors. This is due to our choice of convective parameters, as can be seen in Figure~\ref{fig:fig_4}. Second, the adopted input physics does not allow blue loops to form in models with initial mass less than $4\,\text{M}_{\odot}$, whereas a large fraction of the \cite{Rodriguez2021} sample is comprised of classical Cepheids with implied masses near or below $4$\,M$_{\odot}$, with a wide variety of possible initial rotation rates. Note that a small number of data points is consistent with Cepheids with initial masses greater than $7$\,M$_{\odot}$, making them interesting objects for further study, given the scarcity of well-studied Cepheids at the high-mass end of the distribution.

Phenomena such as overshooting were simplified in this work by assuming that the overshooting parameter is the same in the envelope and in the core. In the future, a calibration of the overshooting parameter in the core should be performed. Besides, physical processes such as pulsation-driven mass loss are not currently implemented in MESA, and this could have an important impact on the PCRs \citep{Neilson2008}. In order to improve the determination of the IS edges, a study of convective parameters that reproduce an IS that covers all the classical Cepheids from the OGLE catalog of variable stars should be performed. A grid covering a wider range in masses and metallicity would help us cover a wider range of periods and PCRs, better matching the empirical data presented by \cite{Rodriguez2021}. In addition, there is a degeneracy in these models since a PCR vs. period curve with a certain mass, metallicity, and initial rotation rate can be overlapped by a curve with a different combination of these parameters. 

\section*{Acknowledgements}

We thank the referee for her/his helpful comments, which led to an improved presentation of our results. Support for this project is provided by the Ministry for the Economy, Development, and Tourism's Millennium Science Initiative through grant ICN12\textunderscore 12009, awarded to the Millennium Institute of Astrophysics (MAS); by Proyecto Basal ACE210002 and FB210003; and by FONDECYT grant \#1171273. The research leading to these results has received funding from the European Research Council (ERC) under the European Union’s Horizon 2020 research and innovation programme (grant agreements No 695099 and No 951549). FEA acknowledge support from the Polish National Science Center grant SONATA BIS 2020/38/E/ST9/00486. COH and AVN acknowledge support from the National Agency for Research and Development (ANID), Scholarship Program Doctorado Nacional, grant number 2018 – 21180315, and 2020\,–\,21201226, respectively.

\section*{Data Availability}

Evolutionary models were computed with the version 11701 of MESA. The required inlists can be found in the following link \url{https://doi.org/10.5281/zenodo.6603666}

\bibliographystyle{mnras.bst}
\bibliography{main.bib}

\bsp
\label{lastpage}
\end{document}